\documentclass[usenatbib]{mn2e}
\usepackage{amssymb,amsmath,graphicx}
\usepackage{longtable,lscape}
\usepackage{appendix}

\newcommand{\oiii}[1]{[{\ensuremath{\mathrm{O}}}\,\textsc{iii}]\:#1}
\newcommand{\oii}[1]{[{\ensuremath{\mathrm{O}}}\,\textsc{ii}]\:#1}
\newcommand{\oi}[1]{[{\ensuremath{\mathrm{O}}}\,\textsc{i}]\:#1}
\newcommand{\nii}[1]{[{\ensuremath{\mathrm{N}}}\,\textsc{ii}]\:#1}
\newcommand{\n}[1]{[{\ensuremath{\mathrm{N}}}\,\textsc{i}]\:#1}
\newcommand{\sii}[1]{[{\ensuremath{\mathrm{S}}}\,\textsc{ii}]\:#1}
\newcommand{\siii}[1]{[{\ensuremath{\mathrm{S}}}\,\textsc{iii}]\:#1}
\newcommand{\cliii}[1]{[{\ensuremath{\mathrm{Cl}}}\,\textsc{iii}]\:#1}
\newcommand{\ar}[1]{[{\ensuremath{\mathrm{Ar}}}\,\textsc{iv}]\:#1}
\newcommand{\ariii}[1]{[{\ensuremath{\mathrm{Ar}}}\,\textsc{iii}]\:#1}

\newcommand{\feiii}[1]{[{\ensuremath{\mathrm{Fe}}}\,\textsc{iii}]\:#1}
\newcommand{\neiii}[1]{[{\ensuremath{\mathrm{Ne}}}\,\textsc{iii}]\:#1}
\newcommand{\heii}[1]{{\ensuremath{\mathrm{He}}}\,\textsc{ii}\:#1}
\newcommand{\hei}[1]{{\ensuremath{\mathrm{He}}}\,\textsc{i}\:#1}
\newcommand{\hii}{{\ensuremath{\mathrm{H}}}\,\textsc{ii}}
\newcommand{\h}{\ensuremath{\mathrm{H}}}
\newcommand{\hb}{\ensuremath{\mathrm{H}\beta}}
\newcommand{\ha}{\ensuremath{\mathrm{H}\alpha}}
\newcommand{\heps}{\ensuremath{\mathrm{H}\epsilon}}
\newcommand{\hg}{\ensuremath{\mathrm{H}\gamma}}
\newcommand{\hd}{\ensuremath{\mathrm{H}\delta}}

\begin{document}
\title{Strongly star forming galaxies in the local Universe 
with nebular He II 4686 emission}
\author[Maryam Shirazi and Jarle Brinchmann]
{Maryam~Shirazi$^1$\thanks{shirazi@strw.leidenuniv.nl} and 
Jarle~Brinchmann$^1$ \\ 
$^1$Leiden Observatory, Leiden University, P.O. Box 9513, 
2300 RA Leiden, The Netherlands}
\date{Accepted 2011 December 7. Received 2011 December 7; in original form 2011 September 15}
\maketitle

\begin{abstract}
  We present a sample of 2865 emission line galaxies with strong
  nebular $\heii{\lambda4686}$ emissions in Sloan Digital Sky Survey
  Data Release 7 and use this sample to investigate the origin of this
  line in star-forming galaxies. We show that star-forming galaxies
  and galaxies dominated by an active galactic nucleus form clearly
  separated branches in the $\heii{\lambda4686}/\hb$ versus
  $\nii{\lambda6584}/\ha$ diagnostic diagram and derive an empirical
  classification scheme which separates the two classes. We also
  present an analysis of the physical properties of 189 star forming
  galaxies with strong $\heii{\lambda4686}$ emissions. These
  star-forming galaxies provide constraints on the hard ionizing
  continuum of massive stars. To make a quantitative comparison with
  observation we use photoionization models and examine how different
  stellar population models affect the predicted $\heii{\lambda4686}$
  emission. We confirm previous findings that the models can predict
  $\heii{\lambda4686}$ emission only for instantaneous bursts of 20\%
  solar metallicity or higher, and only for ages of $\sim 4-5$ Myr,
  the period when the extreme-ultraviolet continuum is dominated by
  emission from Wolf-Rayet stars. We find however that 83 of the
  star-forming galaxies (40\%) in our sample do not have Wolf-Rayet
  features in their spectra despite showing strong nebular
  $\heii{\lambda4686}$ emission.  We discuss possible reasons for this
  and possible mechanisms for the $\heii{\lambda4686}$ emission in
  these galaxies. 
\end{abstract}

\begin{keywords}
Stars: Wolf-Rayet -- Galaxies: evolution -- Galaxies: stellar content 
-- Galaxies: star formation

\end{keywords}


\section{Introduction}
The ionizing continuum of stars at $\lambda < 912$ \AA\ is of major
importance for interpreting emission line observations of galaxies
because many of the strong lines observed in the spectra of galaxies,
such as $\oiii{\lambda5007}$, $\neiii{\lambda3869}$ and
$\heii{\lambda4686}$, have ionization potentials in excess of 13.6
eV. Despite this importance we are severely limited by interstellar
absorption in observing stellar spectra in this spectral window
directly \citep[e.g.][]{Ho93}. Although we can get direct information
at slightly longer wavelengths with space-based UV spectroscopy
\citep[e.g.][]{crow02}, most of our knowledge about the $\lambda <
912$\AA\ region is based on indirect evidence, even for solar
metallicity.

A promising way to indirectly obtain information on the stellar
ionizing continuum is to compare emission line properties (e.g. flux,
equivalent width) to predictions from photoionization codes such as
CLOUDY \citep{Fer} or MAPPINGS III \citep{Allen}.  In practice these
kinds of studies provide modest constraints on stellar atmosphere
models \citep[e.g.][]{crow99}. However where predictions of models
differ significantly, this approach can yield useful information. This
is the approach we will adopt in this work, where we will make use of
the $\heii{\lambda4686}$ nebular emission line to place constraints on
stellar models and in particular on the ionization mechanism for this
line.


The presence of a nebular $\heii{\lambda4686}$ line in the integrated
spectrum of a galaxy indicates the existence of sources of hard
ionizing radiation as the ionization energy for He$^+$ is 54.4 eV
($\lambda \approx 228$ \AA).  This hard radiation can of course be
produced by an active galactic nucleus (AGN), and most sources with
luminous \heii{} emission, in a flux limited sample, are indeed
galaxies with an AGN\footnote{We will here not distinguish between the
  host galaxy and its nuclear power source so will refer to these
  galaxies as AGNs}. However the required hard radiation can also be
provided by stellar sources and $\heii{\lambda4686}$ emission is
frequently seen in \hii-galaxies. The line appears to be associated
with young stellar populations; for instance, \citet{Berg} proposed O\emph{f}
stars as the sources of $\heii{\lambda4686}$ emission in dwarf
galaxies. Subsequent discussion has mostly focused on Wolf-Rayet (WR)
stars, although the distinction between these two classes is rather
blurred \citep[e.g.][]{Gra11}. \citet[][see also \citet{Sch98}]{Sch96}
showed that the hard radiation field of WR stars could provide a good
explanation of the nebular $\heii\lambda4686$ seen in \hii-galaxies.
\cite{Gu00} tried to test this in a careful study of \hii-galaxies
with prominent WR features. They were however unable to find WR
features in 12 out of the 30 galaxies with nebular $\heii{\lambda
  4686}$ emission. The same lack of WR features in metal poor Blue
Compact Dwarf (BCD) galaxies was pointed out by \citet{Thu05}. They
proposed that fast radiative shocks could be responsible for this
emission (see also \cite{Gar}).

Similar results were reported by \citet[hereafter B08]{Br08}, who
analysed a sample of strong emission line galaxies in the Sloan Digital Sky Survey \citep{Yo} with
$\heii{\lambda4686}$ emission. They showed that at least at
metallicities of $12+\log \mathrm{O/H}>8$, there appeared to be a
close correlation between WR features in galaxies and the presence of
$\heii{\lambda 4686}$ emission, but this appeared not to be so
clear-cut at lower metallicities.

This apparent lack of connection of \heii{} emission with the hard UV
radiation from WR stars has also been seen in spatially resolved
spectroscopic studies. \citet{Keh08} performed an integral field
spectroscopy study for the \hii\ galaxy II\,Zw\,70 and found that the
region associated with nebular $\heii{\lambda4686}$ emission
was a few arcsec offset from the region with detected WR
features. More recently \citet{Keh11} and \citet{NM} have presented
studies of $\heii{\lambda 4686}$ emission in M33. Both studies find
some regions with nebular $\heii{\lambda 4686}$ emission that are not
associated with WR stars (see also \citet{Had07}, \citet{Lopez} and \citet{Moreal}). 

Thus a series of studies have shown that while $\heii{\lambda4686}$
emission frequently is found in association with WR stars, it
appears not to be so in all cases, particularly at low
metallicity. As mentioned above, possible additional sources
of high energy photons could be X-ray binaries \citep{Gar},
strong shocks \citep{Dop96}, low-level AGN activity and alternative
models for stellar evolution \citep{Yoon}. However the existing
studies do not show clear trends that allow us to distinguish between
these scenarios.

Crucially the samples in most of the previous studies have
not been selected specifically to study \heii{} emission lines. To
make progress in understanding this puzzle it is important to have as
large as possible sample of \heii{} emitting galaxies to allow one to
study the relationship between \heii{} emission and other physical
properties. To this end we present here an analysis of emission line
galaxies with strong $\heii{\lambda4686}$ emission in SDSS 
Data Release 7 \citep[DR7,][]{Ab}.
 
In section~\ref{sec:data} we discuss the sample selection and
carefully account for AGN emission. The physical properties of the
\heii{} emitting galaxies are discussed in section~\ref{sec:physical}
and the observed $\heii{\lambda 4686}/\hb$ ratios are compared to
model predictions in section~\ref{sec:prediction}.  In
section~\ref{sec:origin} we test these model predictions and investigate whether the
presence of $\heii{\lambda 4686}$ is associated with WR features. We
find that low metallicity systems frequently do not show signs of WR
stars. We discuss possible explanations for this finding in
section~\ref{sec:explanation} and conclude in
section~\ref{sec:conclusion}.

\section{Data}
\label{sec:data}
Our sample is based on galaxy spectra from SDSS DR7 which cover a 
wavelength range of 3800-9200 \AA. The spectra were analysed using
the methodology discussed in \citet{Tr} (see also Brinchmann et al. 2004)
to provide accurate continuum subtraction. 
All emission line sources were additionally analysed using the pipeline 
discussed in B08 to measure a wider gamut of emission lines. For each 
galaxy we measure 40 lines, These lines and the number of spectra that 
show these lines with $S/N > 5.5$ are summarised in Table \ref{table0}.

\begin{table}
   \centering
   \begin{tabular}{|l||rr|}
      \multicolumn{3}{l}{} \\
      \hline
      Measured line  & Number of spectra & Fraction\\
      \hline\hline
    $\oii{\lambda 3726,3729}$  & 243977 & 16.51\% \\
      $\neiii{\lambda 3869}$  &  39143 &  2.65\% \\
      $\h{8}$  &  92230 &  6.24\% \\
      $\neiii{\lambda 3967}$  &  18367 &  1.24\% \\
      $\heps$  &  64120 &  4.34\% \\
      $\hei{\lambda 4026}$  &   6368 &  0.43\% \\
      $\sii{\lambda 4069}$  &   1503 &  0.10\% \\
      $\hd$  & 163534 & 11.07\% \\
      $\hg$  & 295528 & 20.00\% \\
      $\oiii{\lambda 4363}$  &   7886 &  0.53\% \\
      $\hei{\lambda 4472}$  &  15563 &  1.05\% \\
      $\feiii{\lambda 4658}$  &   2626 &  0.18\% \\
      $\heii{\lambda 4685}$  &   4034 &  0.27\% \\
      $\ar{\lambda 4711}$  &   2893 &  0.20\% \\
      $\ar{\lambda 4740}$  &   1074 &  0.07\% \\
      $\hb $  & 458324 & 31.02\% \\
      $\oiii{\lambda 4959}$  & 147734 & 10.00\% \\
      $\oiii{\lambda 5007}$  & 339212 & 22.96\% \\
      $\n{\lambda 5197}$  &   1318 &  0.09\% \\
      $\n{\lambda 5200}$  &    758 &  0.05\% \\
      $\cliii{\lambda 5518}$  &    160 &  0.01\% \\
      $\cliii{\lambda 5538}$  &    190 &  0.01\% \\
      $\nii{\lambda 5755}$  &   4888 &  0.33\% \\
      $\hei{\lambda 5876}$  &  75256 &  5.09\% \\
      $\oi{\lambda 6300}$  & 119328 &  8.08\% \\
      $\siii{\lambda 6312}$  &   5552 &  0.38\% \\
      $\oi{\lambda 6363}$  &  16100 &  1.09\% \\
      $\nii{\lambda 6548}$  & 361641 & 24.48\% \\
      $\ha$  & 613338 & 41.51\% \\
      $\nii{\lambda 6584}$  & 562659 & 38.08\% \\
      $\hei{\lambda 6678}$  &  19084 &  1.29\% \\
      $\sii{\lambda 6717}$  & 410903 & 27.81\% \\
      $\sii{\lambda 6731}$  & 341386 & 23.11\% \\
      $\hei{\lambda 7065}$  &   4232 &  0.29\% \\
      $\ariii{\lambda 7135}$  &  24034 &  1.63\% \\
      $\oii{\lambda 7318,19,29,30}$  & 659 &  0.04\% \\
      \hline
   \end{tabular}
   \caption{The table shows number of spectra that have the indicated line detected at  
    $S/N>5.5$. The total number of analysed spectra is 1,477,411.}
   \label{table0}
\end{table}

Our concern in this paper is not to analyse a volume- or
magnitude-limited sample of galaxies, we therefore do not impose a
redshift cut nor a magnitude limit. Since the blue wavelength cut-off
of the SDSS spectrograph is $\sim 3800$\AA, the
$\oii{\lambda3727,3729}$ doublet falls outside the spectral range for
$z < 0.02$. This is a concern, because as we will see later 55\% of
our final sample fall in this region of redshift space and their
oxygen abundances are therefore somewhat uncertain. When possible we
use the $\oii{\lambda7318-7330}$ quadruplet instead \citep{Knia04},
but as this is a fairly weak line and falls in a region with
significant sky emission, we cannot always make use of this line.

\subsection{Sample selection and classification}
\label{sec:sample}
We select our sample requiring signal to noise ratio $>5.5$ in
$\heii{\lambda4686}$, the resulting data set is given in Table \ref{table5}
\footnote{The full table of 3292 spectra is available in electronic form 
in http://www.strw.leidenuniv.nl/$\sim$shirazi/SB011/.}. When the width of the \heii{} line is consistent
with that of the strong forbidden lines, we make the assumption that
it has a nebular origin. Given that \heii{} lines from individual WR
stars typically are considerably broader than the forbidden lines in
galaxies (e.g.\ B08), we feel this is a reasonable assumption. In
addition we require a S/N$>3$ in each of \hb, $\oiii{\lambda5007}$,
\ha\ and $\nii{\lambda6584}$ emission lines to reliably classify our galaxies
(Brinchmann et al.\ 2004, hereafter B04)\nocite{B04}. The resulting sample contains 2865 spectra with strong
nebular $\heii{\lambda4686}$ emission. 
\setcounter{table}{2}
In parallel, the spectra of the sample galaxies are examined for WR
signatures using the approach discussed by B08.  This resulted in a
total sample of 385 spectra with likely and secure WR features (Class
1, 2 and 3 from B08). While we do not discuss the sample of all WR
galaxies in DR7 with Class 1--3 in detail here, we note that it intersects
that of the \heii{} sample but is not a strict subset of it (see
Table~\ref{table1}).

\begin{figure}
\centerline{\hbox{\includegraphics[width=0.35\textwidth, angle=90]
             {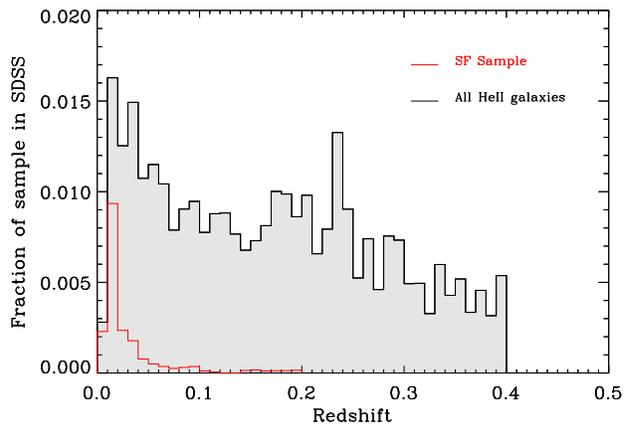}}}
         \caption{The redshift distribution of the full $\heii{}$
           sample, relative to that of the parent sample is shown as a
           shaded grey histogram.  The red histogram shows the
           redshift distribution of the $\heii{}$ SF sample relative
           to that of all star-forming galaxies in the parent sample;
           the parent sample consisting of those galaxies in the SDSS
           DR7 having $\hb$, $\oiii{\lambda5007}$, $\ha$ and
           $\nii{\lambda6584}$ detected at S/N$ > 3$. In comparison
           with the SDSS, it is clear that the galaxies in our SF
           sample have a redshift distribution strongly shifted toward
           low redshift.}
 \label{fig1}
\end{figure}

Figure~\ref{fig1} shows the redshift distribution for the fraction of
the \heii{} sample in the SDSS as a shaded grey histogram. The cut-off
at $z\sim 0.4$ is due to \ha\ falling outside the spectrograph
range. The red histogram shows the redshift distribution for the
fraction of just the \emph{star-forming} (SF) galaxies in the SDSS
showing $\heii{\lambda4686}$ emission (see below for a discussion of
the classification).  For each class we have divided the number of
\heii{} emitting galaxies in that redshift bin by the number of
similarly classified galaxies in the parent sample (SDSS galaxies that
have S/N$>3$ in each of \hb, $\oiii{\lambda5007}$, \ha\ and
$\nii{\lambda6584}$) in that redshift bin. A constant value would
therefore indicate a similar redshift distribution of the \heii\
sample and the parent sample. It is clear from this that full \heii\
sample closely follows the overall distribution of the SDSS, but that
the star-forming galaxies with $\heii{\lambda4686}$ emission are
predominantly found at low redshift. We can also see that
less than 2\% of all galaxies, and less than 1\% of the star-forming
galaxies in the SDSS DR7 show $\heii{\lambda4686}$ emission in their
spectra.

\begin{figure}
\centerline{\hbox{\includegraphics[width=0.35\textwidth, angle=90]
             {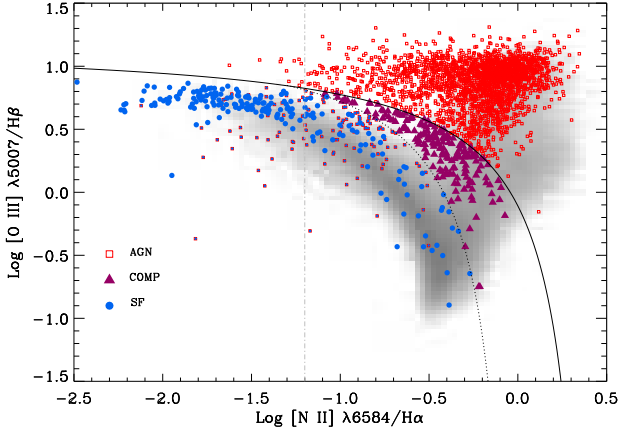}}}
         \caption{This plot shows the BPT diagnostic diagram for the
           sample.  The Ka03 classification line is shown as a dotted
           line and Ke01 classification line as a solid line. The grey
           dashed-dotted line shows the N2 PP04 metallicity
           calibration for $\rm{12+\log O/H}=8.2$. The distribution of
           emission line galaxies in the SDSS is shown by the
           gray-scale 2D distribution where the grey-scale shows the
           logarithm of the number of galaxies in each bin.  Blue
           circles show star-forming galaxies, triangles show
           composite galaxies and red squares mark AGNs. As discussed
           in the text, for some galaxies the classification has been
           adjusted which is why some objects in the star-forming
           region are classified as AGN. We mark these galaxies with a
           blue plus over the red square.}
\label{fig2}
\end{figure}

To classify the dominant ionization source in each galaxy we follow
previous studies in using the \citet[][BPT]{BPT} line ratio diagnostic
diagram of $\oiii{\lambda5007}/\hb$ versus $\nii{\lambda6584}/\ha$ as
our starting point (Figure~\ref{fig2}). As has been discussed
extensively \citep[e.g.][]{Ter91,Ke01,Ka03,Ke06,Stas06} this diagram
allows a separation of AGN and star-forming galaxies because of their
significantly different ionizing spectra, typically leading to high
$\oiii{\lambda5007}/\hb$ and $\nii{\lambda6584}/\ha$ when an AGN is
dominating the output of ionizing photons.  Ke01 used a combination of
stellar population synthesis and photoionization models to compute a
theoretical maximum starburst line that isolates objects whose
emission line ratios can be accounted for by photoionization by
massive stars (below and to the left of the curve) from those where
some other source of ionization is required. Ka03 defined an empirical
upper limit to the \hii\ region sequence of SDSS galaxies in the BPT
diagram. The region lying between these two lines represents objects
more naturally explained as having a composite spectrum combining
\hii\ region emission with a harder ionizing source. As we will see
later, this interpretation is further corroborated by the
$\heii{\lambda 4686}/\hb$ ratios we find for our sample.

In the present study, we adopt a two-stage classification
methodology. We start out by classifying all galaxies using the BPT
diagram, and we will then refine our classifications for a subset of
the galaxies using detailed inspection of the spectra and the
$\heii{\lambda 4686}$ line properties. For the initial classification
we adopt a similar methodology to that of B04 and use the separation
criteria defined by Ke01 and Ka03 to divide galaxies into different
classes. Galaxies which are distributed above the Ke01 dividing line
are considered AGNs, Galaxies between the Ke01 and Ka03 limits have
composite classification, which means their source of ionizing
radiation could be a combination of star formation and AGN activity.
Finally, galaxies below the Ka03 line have a SF classification, which
as we will see might be modified subsequently.

Figure~\ref{fig2} shows the BPT diagram for our sample, the Ke01 and
Ka03 classification lines are shown as solid and dotted lines,
respectively. The distribution of all emission line galaxies in the
SDSS with $S/N>3$ in \hb, $\oiii{\lambda5007}$,
\ha\ and $\nii{\lambda6584}$ is shown as a grey-scale 2D distribution,
where the grey-scale shows the logarithm of the 
number of galaxies in each bin. Blue circles show star-forming
galaxies, triangles show composite galaxies and red
squares mark AGNs. The grey dashed-dotted line shows the N2 \citep[hereafter PP04]{PP04} 
metallicity calibration for $\rm{12+\log O/H}=8.2$ ($\nii{\lambda6584}/\ha=-1.2$). 
These classifications are the final ones and
incorporate further information as described in the following.

While the BPT diagram is a useful classification diagram, it is not
particularly sensitive to low levels of AGN contamination and some
progress can be made by including lines originating in the mostly
neutral ISM \citep[e.g.][]{Ke06}. For our purposes we however need to
be very confident in the lack of AGNs in our sample and we will use
the $\heii{\lambda4686}/\hb$ ratio for this purpose.

\begin{figure}
\centerline{\hbox{\includegraphics[width=0.35\textwidth, angle=90]
             {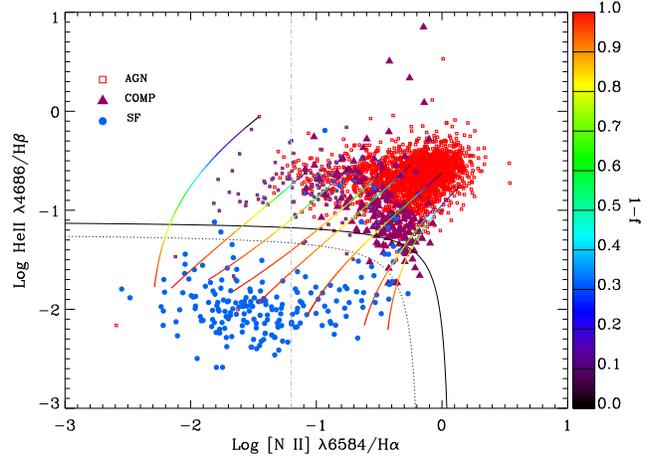}} }
         \caption{This plot shows our sample in the
           $\heii{\lambda4686}/\hb$ versus $\nii{\lambda6584}/\ha$
           diagnostic diagram. The dotted line shows an empirical line
           separating AGN and composite objects from star-forming
           galaxies. The grey dashed-dotted line shows the N2 PP04
           metallicity calibration for $\rm{12+\log O/H}=8.2$.
           Symbols are the same as the BPT diagram. As
           $\heii{\lambda4686}$ have a higher ionization potential in
           comparison to $\oiii{\lambda5007}$ and is much less
           sensitive to the electron temperature, we can clearly see
           separation between the classes in this diagram.  The lines
           with colour gradients in the figure are simulated fluxes
           drawn from adding a random set of star-forming galaxy from
           the SDSS with emission line fluxes of a random set of AGN
           spectrum.  The colouring of the lines corresponds to the
           fraction of the spectrum contributed by the star-forming
           galaxy, $1-f$, as indicated by the colour bar on the
           side. The solid line shows the theoretical upper limit for
           $\heii{\lambda4686}/\hb$ ratio.  } \label{fig3}
\end{figure}

As remarked earlier, only photons with energy in excess of 54.4 eV can
ionize He$^{+}$ and thence produce the $\heii{\lambda 4686}$
recombination line. If we consider single stellar population models
from Starburst99 \citep{Le} at an age of 2 Myr (so that all stars are on the
main sequence), we find that the integrated spectrum of this
population typically contain 2-3 orders of magnitude fewer photons at
this energy than at the energy required to ionize O$^{+}$ (35.5 eV),
which is needed to produce the $\oiii{\lambda5007}$ line from
collisionally excited O$^{+2}$.
This line is therefore a very sensitive probe of AGN activity,
particularly if used in conjunction with other line ratios. We
therefore make use of the $\heii{\lambda4686}/\hb$ vs
$\nii{\lambda6584}/\ha$ diagram to further refine the classification
of our sources.  We show this diagram in Figure~\ref{fig3} where the
symbols and colours are the same as in Figure~\ref{fig2}.
Note that there is generally a very significant offset between the
star-forming galaxies and AGNs in this diagram, in contrast to the
gradual transition in much of the BPT diagram. To make a quantitative
separation, we draw a random set of star-forming galaxies from the
SDSS and gradually add their emission line fluxes to that of a random
set of AGNs. We quantify this by the variable $f$ which is defined to
be the fraction of the total \hb\ flux comes from the AGN.  The total
flux is therefore $f \times AGN_{\textsc{Flux}} + (1-f) \times
SF_{\textsc{Flux}}$.  Changing $f$ will trace out a path in the
diagnostic diagram in Figure~\ref{fig3} as illustrated by the lines
with colour gradients in the figure. The colouring of the lines
corresponds to $1-f$, as indicated by the colour bar on the
side. We repeat this process for a thousand AGN and SF objects
located in different bins in the \heii/\hb\ diagram.

Based on this analysis we find that there is a well defined locus
where 10\% percent of the $\heii{\lambda4686}$ flux comes from an AGN
(in this case $\approx 1$\% of the \hb\ flux comes from the
AGN, see section \ref{sec:AGN} for further details). A good
fit to this relation is given by:
\begin{equation}
\label{eq:CF0}
\log (\frac{F_{\heii{\lambda{4686}}}}{F_{\hb}}) = 
-1.22+
\frac{1}{8.92 \log (\frac{F_{\nii{\lambda{6584}}}}{F_{\ha}}) + 1.32},
\end{equation}      
which is shown as a dotted line in Figure~\ref{fig3}. The
solid line shows a theoretical maximum starburst line, similar to the Ke01 line
in the BPT diagram --- we discuss this further below.

Finally, we look at the spectra of galaxies that we would classify as
star-forming on the basis of their location in the BPT diagram, but
that are offset from the rest of the star-forming galaxies in the
\heii/\hb\ diagram and check whether they show AGN features such as
broad Balmer lines, strong $\mathrm{Ne\textsc{v}}\lambda3426$,
Fe\,\textsc{ii} emission or if they show a similar
$\oiii{\lambda4363}/\hg$ ratio to that of AGNs. If some of these
features are present, we change the classification from star-forming
to AGN (these objects are plotted as red squares containing a blue
cross in Figures ~\ref{fig2} and~\ref{fig3}). This happens for 127
(39\%) galaxies. This is a conservative approach as we would exclude
star-forming galaxies with strong outflows and hence a broad base to
the Balmer lines for instance.

To estimate the maximum starburst line in Figure~\ref{fig3}, we
adopted the Charlot \& Longhetti (2001, hereafter CL01)\nocite{CL01}
models. These combine evolving stellar populations models from Bruzual
\& Charlot (unpublished BC00 models using Padova 1994 tracks, see
Bruzual \& Charlot 2003 (BC03) for the current models)\nocite{BC03} with the
photoionization code Cloudy \citep{Fer} and adopt the simple dust
attenuation prescription of \citet{CF00}.  The main model parameters
for our calculations are the metallicity $Z$, the ionization
parameter, $U$, the dust attenuation $\tau_{\mathrm{V}}$ and the
dust-to-metal ratio, $\xi$, the model parameters used are given in
Table \ref{table2}, see B04 for a more detailed discussion.
We use the CL01 Single Stellar Population (SSP) models since these
achieve the highest possible $\heii{\lambda4686}/\hb$ values. We then
identify the maximum ratio reached by the different models and use
this upper envelope to define the maximum starburst line shown as a
solid line in Figure~\ref{fig3}.


%
%
%

\begin{table}
\label{tab:summary}

\begin{center}
\begin{tabular}{|l|l|c|}
 \hline
\multicolumn{1}{l}{Sample} & \multicolumn{1}{l}{He\,\textsc{ii}} &
\multicolumn{1}{l}{He\,\textsc{ii} + WR} \\
\hline
\hline
  Total & 2865 & 385 \\
  AGN & 2474 & 234 \\
  Star forming & 199 & 116 \\
  Composite & 179 & 35 \\
\hline
\end{tabular}
\end{center}
\caption{An overall numerical summary of the \heii{} sample. 
See section~\ref{sec:data} for a summary of the selection and
classification details.  See section~\ref{sec:origin} for details on
the WR classification.}
\label{table1}
\end{table}
This combination of classification methods means that there is not a
one-to-one mapping between the location of an object in a diagnostic
diagram and its final classification, as is clear from
Figure~\ref{fig2} and~\ref{fig3}.  We also mention that we do not
change classes for galaxies classified as AGNs or composites using the
BPT diagram, but which fall within the star-forming region in the
\heii{}/\hb\ diagram and this is the reason why there are 7 AGN below
our star-formation--AGN dividing line.

By contrasting Figure~\ref{fig2} and~\ref{fig3} we can make a couple
of interesting observations. The first is that while in the BPT
diagram we see a steady increase in $\oiii{\lambda5007}/\hb$ with
decreasing $\nii{\lambda6584}/\ha$, in Figure~\ref{fig3} we see no
major change in $\heii{\lambda4686}/\hb$ with
$\nii{\lambda6584}/\ha$. 
This indicates that $\heii{\lambda4686}/\hb$ and consequently the
ionizing spectrum of stars at $\lambda<228$ \AA\ vary only weakly with
metallicity. The second difference between the two plots is that the
$\heii/\hb$ diagram can separate star-forming galaxies from composite
galaxies better than BPT diagram since the $\heii{\lambda4686}/\hb$
ratio is more sensitive to the hardness of the ionizing source than
$\oiii{\lambda5007}/\hb$. We can use this to further support our
supposition that the gas in the galaxies falling between the Ka03 and
Ke01 lines in the BPT diagrams is ionized by a combination of stars and
an AGN --- while they are adjacent to the star-forming sequence in the
BPT diagram they nearly all clearly separate from the star-forming
sequence in the $\heii{}$/\hb\ diagram, corresponding to an AGN
contribution to $\heii{\lambda 4686}>50$\%. Thus referring to these as
composite objects appear to be justified.

\begin{figure}
\centerline{\hbox{\includegraphics[width=0.35\textwidth, angle=90]
             {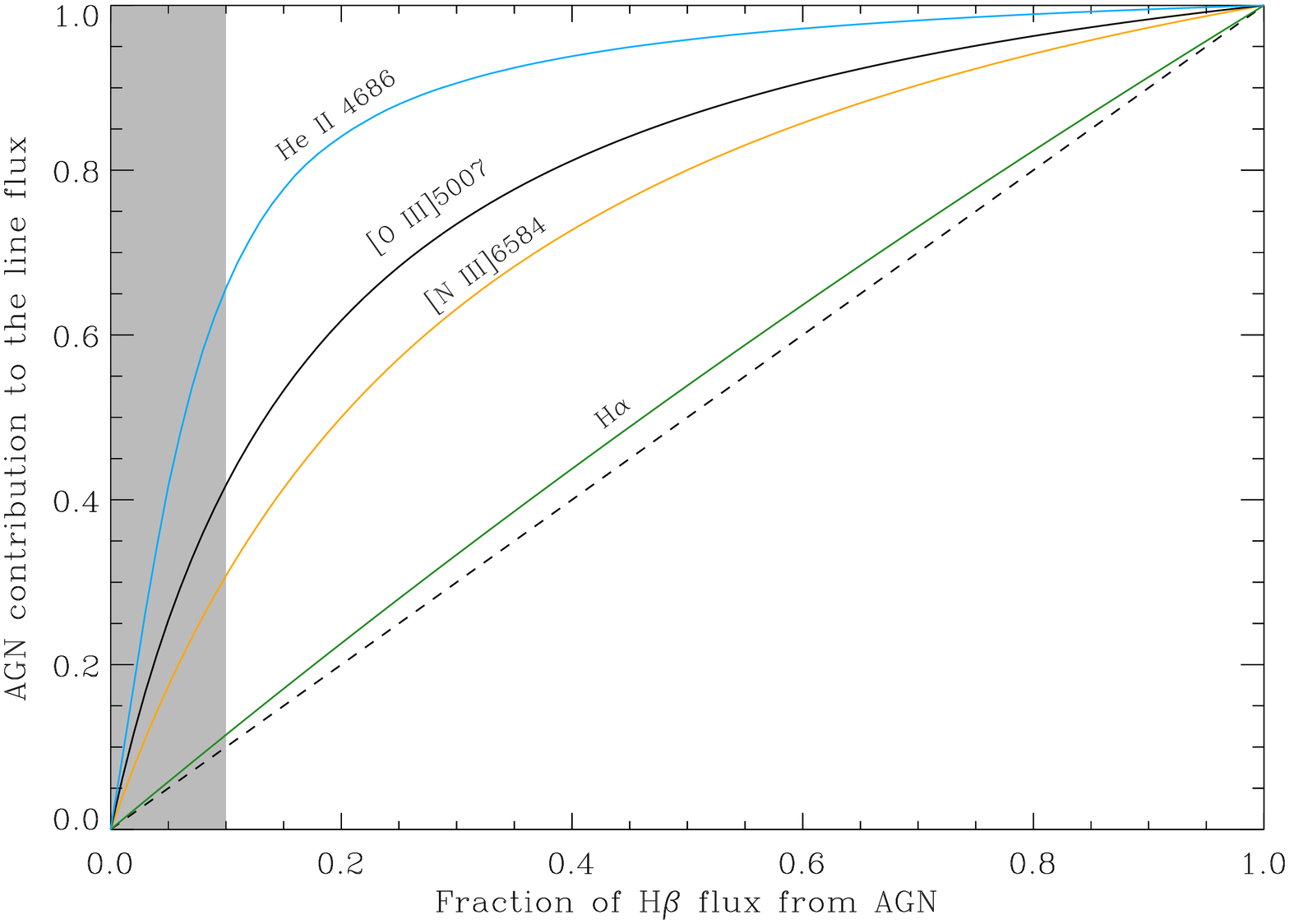}} }
\centerline{\hbox{\includegraphics[width=0.31\textwidth, angle=90]
             {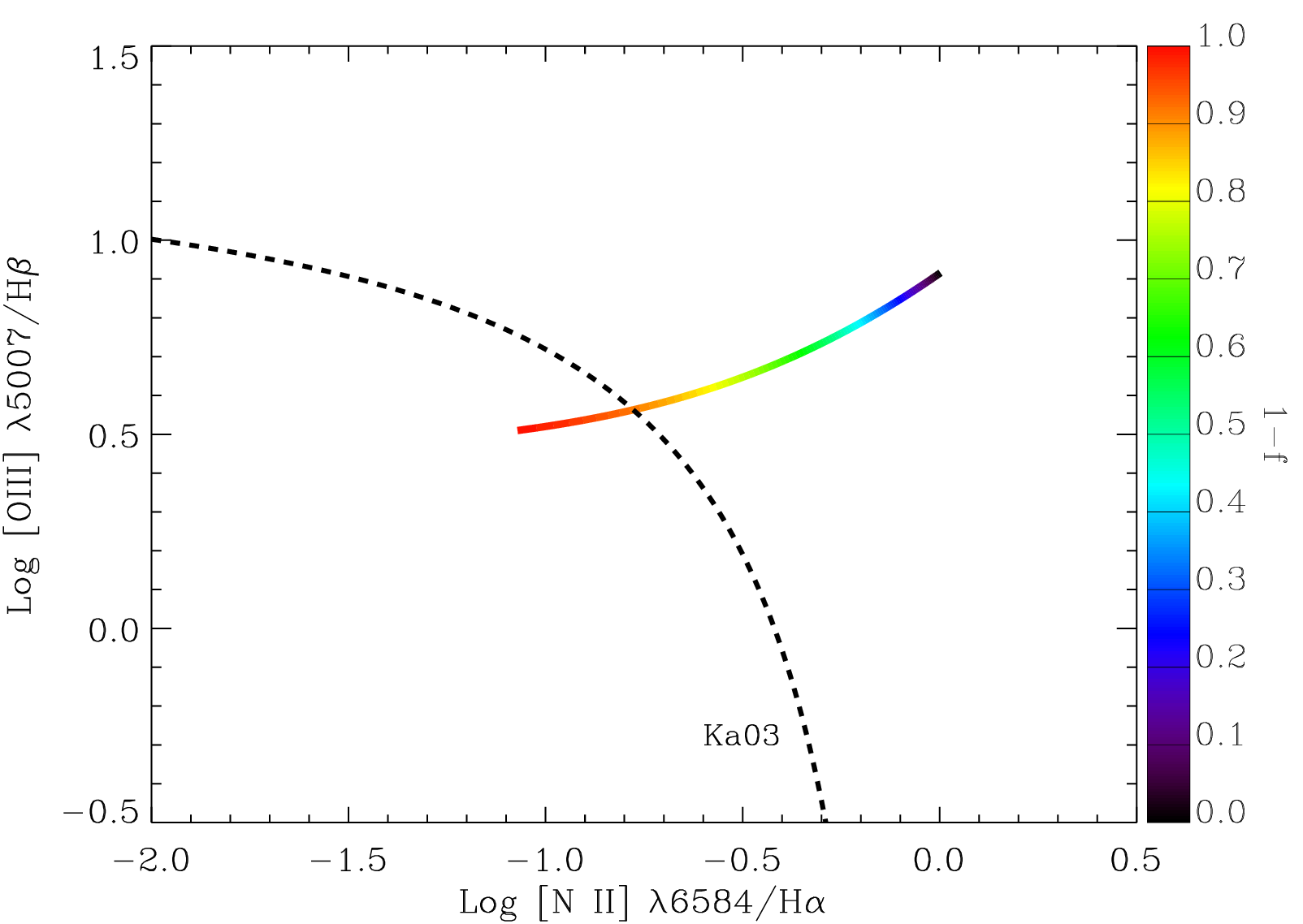}} }
         \caption{The upper panel shows the sensitivity of different
           lines to the presence of an AGN as a function of the
           fraction of the \hb\ flux originating from an AGN. The grey
           shading indicates the region where a typical galaxy would
           be classified as star-forming in the BPT diagram (see text
           for details). We see that we might classify galaxies as
           star-forming even if 10\% of the \hb\ flux comes from AGN.
           The bottom panel illustrates the method used and shows the
           path traced by a galaxy in the BPT diagram as an increasing
           amount of AGN light is added to a star-forming galaxy
           spectrum. The colouring of the line corresponds to the
           fraction of the \hb\ line flux contributed by the
           star-forming galaxy, $1-f$, as indicated by the colour bar
           on the side. The dashed line is the Ka03 classification
           line, the intersection of this line with the trajectory of
           the simulated flux is at $f\approx 0.1$.}
\label{fig5}
\end{figure}

\subsection{AGN contamination estimation}
\label{sec:AGN}
In view of the clear separation of the AGN and star-formation branches
in Figure~\ref{fig3}, and our sensitivity to low-level AGN
contamination, it is beneficial to study the impact of an AGN on the
line ratios in more detail. This has been discussed in previous
studies (e.g. B04, \citet{Stas06}) but here we extend those efforts to
include $\heii{\lambda 4686}$.

We focus our attention on the BPT diagram as it is most widely used
for emission line classification. We follow the same approach as in
the previous section of adding gradually more of an AGN emission line
spectrum with $\heii{\lambda 4686}$ in emission at S/N$>3$, to a
star-forming one, and find where it intersects the Ka03 line (see
bottom panel of Figure 4). At this point the galaxy would cease to be
classified as a star-forming galaxy.  We repeat this for a total of
10,000 random combinations of spectra. The median AGN contribution to
\hb\ at the point where a galaxy ceases to be classified as
star-forming is $\approx 10$\%.

Figure 4 shows the region where a median galaxy would be classified as
star-forming as the gray shaded region. On top of this we show the
median trend for the fraction of flux in the indicated line as a
function of the fraction of the \hb\ flux coming from an AGN (shown
for reference as the dashed diagonal line). We note that the exact
shape and location of the $\nii{\lambda6584}$ and $\oiii{\lambda5007}$ lines does depend
somewhat on the AGN sample chosen but the qualitative trend remains
the same.

What this figure shows, is firstly the well-known result that some
lines are more sensitive to the presence of an AGN than others. As
mentioned before, the Balmer lines are expected to have less than
$10$\% contribution from an AGN, while the $\nii{\lambda{6584}}$ line can have
more than 30\% of its flux coming from an AGN, putting in question its
use as an abundance indicator on its own (see also
\citet{Stas06}). But for our purposes, it is more important to note
that by adopting a classification based on the BPT diagram, we would
classify a galaxy as star-forming even when $\sim 65$\% of its
$\heii{\lambda 4686}$ flux would come from an AGN.

We can now combine this with our previous result in Figure~\ref{fig3},
where we found that only 10\% of the $\heii{\lambda{4686}}$ emission
comes from an AGN in our refined star-forming sample. Applying this to
Figure~\ref{fig5}, we conclude that less than 1\% of the \hb\ flux
comes from an AGN and the other lines will also only have very small
contributions from an AGN implying we have a quite pure star-forming
sample.

Our final sample of \heii\ star-forming galaxies consists of 189
star-forming galaxies (199 spectra which have been summarised in Table \ref{table4}).  
As mentioned above we have also checked these spectra for the
presence of WR signatures. We will return to a detailed discussion of
this in section~\ref{sec:origin} but will use the result of this
classification in the following plots.

\begin{table}
\label{tab:summary2}

\begin{center}
\begin{tabular}{|l|l|}
 \hline
\multicolumn{1}{l}{Parameter}  &
\multicolumn{1}{l}{Range} \\
\hline
\hline
$Z$,  The metallicity & $-1<logZ/Z_{\odot}<0.6$, 24 steps \\
$U$,  The ionization parameter & $-4.0<log U< -2.0$, 33 steps \\
$\tau_{V}$, The total dust attenuation & $0.01<\tau_{V}<4.0$, 24 steps \\
$\xi$,   The dust-to-metal ratio & $0.1<\xi<0.5$, 9 steps \\
\hline

\end{tabular}
\end{center}
\caption{The model grid used for the present work. We calculate this
  both for a constant star formation history at $t=10^{8}$ yrs as well
  as for an SSP.}
\label{table2}
\end{table}
\section{Physical properties of the sample}
\label{sec:physical}
While the majority of the galaxies in our sample have some physical
parameters in the MPA-JHU value added
catalogues\footnote{\texttt{http://www.mpa-garching.mpg.de/SDSS/DR7}},
our galaxies are sufficiently extreme that we need to rederive some
properties and add some physical parameters to what is in the MPA-JHU
catalogue.

For the calculation of physical parameters we will adopt the Bayesian
methodology outlined by Ka03 and B04. For each model we calculate the
probability of that model given the data assuming Gaussian noise and
obtain the Probability Distribution Function (PDF) of every parameter of
interest by marginalisation over all other parameters (see Appendix
A). We take the median value of each PDF to be the best estimate
of a given parameter.
\begin{figure}
\centerline{\hbox{\includegraphics[width=0.35\textwidth, angle=90]
             {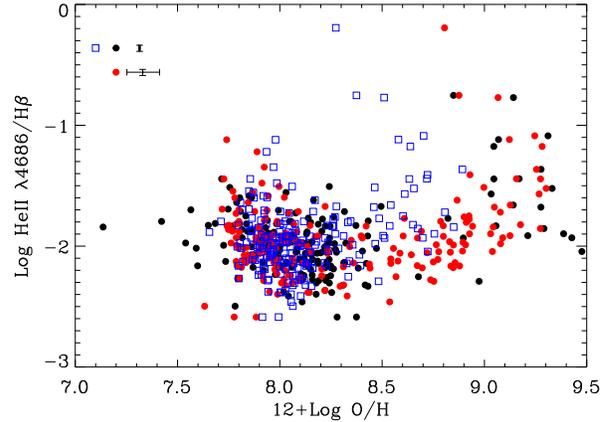}}}
         \caption{This plot shows $\heii{\lambda{4686}}/\hb$ as a
           function of oxygen abundance (see text for details on the
           calculations). Oxygen abundances derived using the direct
           method are shown by black circles, we can not calculate
           oxygen abundances for 13 objects with this method as no
           $\oii$ lines are available for them. Oxygen abundances
           derived from fits to the CL01 models are shown by red
           circles while the blue squares show the O3N2 oxygen
           abundance estimates. Note that the qualitative trends are
           similar, but the O3N2 estimator does not reach as high O/H
           values as the other two models. At low metallicity the
           three methods are in good agreement.}
\label{fig77}
\end{figure}
\subsection{Mass measurements}
\label{sec:abundance}

The standard SDSS pipeline often segment nearby actively star-forming
galaxies incorrectly, thus we need to redo the photometry of our
galaxies. We do this using the Graphical Astronomy and Image Analysis
Tools
(GAIA\footnote{\texttt{http://astro.dur.ac.uk/\%7Epdraper/gaia/gaia.html}}). Most
of our galaxies have strong emission lines within some of the
broad-band filters. Prior to fitting we therefore correct the
magnitudes for emission line contributions by assuming that the
relative contribution of the lines found in the SDSS fiber spectra is
applicable to the galaxy as a whole. As most galaxies in our sample
appear to have a uniformly blue colour, presumably due to active star
formation, we expect this to be a reasonable assumption.

Stellar masses are calculated as outlined above, by fitting a large
grid of stochastic models to the SDSS u, g, r, i, z band
photometry. The grid contains pre-calculated spectra for a set of
100,000 different star formation histories using the BC03 population
synthesis models, following the precepts of \citep{Gal05,Gal08}.

\subsection{Emission line derived parameters}
\label{sec:derived}

We use the CL01 model to analyse the emission lines in our sample. We
adopt a constant star formation history (SFH) and use the same grid
used by B04 (see Appendix A and B04 for further details). In total the
model grid used for the fits have $2\times10^5$ different models.

%
\begin{figure}
\centerline{{\includegraphics[width=0.35\textwidth, angle=90]
           {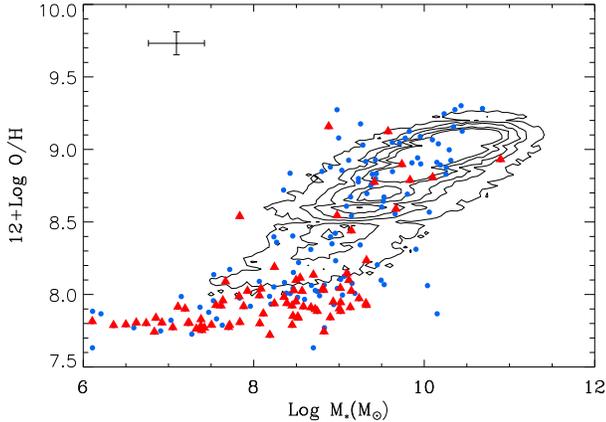}}} 
\caption{ The contours show the mass-metallicity relation for SDSS
  galaxies \citep{Tr}. The present sample of star-forming galaxies
  with WR features is shown 
by blue circles, while the red triangles show the locations of those
that do not show WR features. At low mass we see that our sample is
offset from the rest bulk of the of SDSS but overall they sample much
the same region.}
\label{fig6}
\end{figure}
The main quantity of interest for the present discussion is the oxygen
abundance, quantified as $12+\log \rm{O/H}$.  As there are significant
differences between methods for estimating oxygen abundance
\citep{Ke08}, we have complemented the estimate from the CL01 method
with two independent methods: Firstly, we estimate gas-phase oxygen
abundances with the empirically calibrated estimators proposed by
PP04. They used the line ratios of
$\oiii{\lambda5007}/\hb/\nii{\lambda6584}/\ha$, the O3N2 method, and
$\nii{\lambda6584}/\ha$, the N2 method, as abundance indicators. For
all objects with detected $\oiii{\lambda4363}$ with $S/N>3$, we also
use the $T_{e}$ method, or direct method, using the fitting formulae
provided by \citet{Iz06} to estimate the oxygen abundances.
For those objects without $\oiii{\lambda4363}$, we adopt the
electron temperature estimates from the CL01 models fits for the
direct method calculation.  Whenever we do not have
$\oii{\lambda3727,3729}$, we use the $\oii{\lambda7320,7330}$ lines to
calculate abundances.  However, for 13 objects we are unable to use
the direct method for estimating oxygen abundance as none of the
$\oii$ lines are available.

Figure~\ref{fig77} compares different abundance indicators in the
$\heii{\lambda{4686}}/\hb$ flux ratio versus oxygen abundance
plane. The oxygen abundances derived using the direct method are shown
by black circles while those derived from the fit to the CL01 model
are shown by red circles, the blue squares show O3N2 oxygen
abundances.  The main conclusion we can draw from this comparison is
that all methods agree well at low metallicity while at high
metallicity the trends are similar but the O3N2 estimator reaches a
lower maximum O/H. For concreteness we will adopt the CL01 estimates
for the remainder of the paper but as our main focus will be on the
low metallicity region, our results are robust to the estimator chosen.

Figure \ref{fig6} shows the mass-metallicity relation for the sample
compared to the mass-metallicity relation for all star-forming SDSS
galaxies \citep{Tr} shown as a contour. The sample galaxies with WR
features in the spectra are shown as blue circles, while those that do
not show WR features are plotted as red triangles, we will use the
same symbols in the following. Overall there is a reasonable agreement
with the main SDSS sample except for an offset towards slightly lower
metallicity at a fixed mass at low masses.
\begin{figure}
\centerline{\includegraphics[width=0.35\textwidth, angle=90]
       {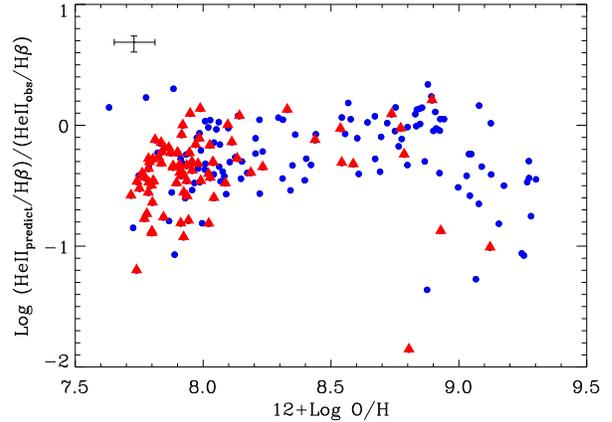}} 
     \caption{This figure shows the logarithm of the ratio of the
       model prediction for $\heii{\lambda4686}/\hb$ to the observed
       ratio as a function of oxygen abundance.  Red triangles show
       the ratio for objects without WR features.  It is clear that
       there is good agreement in the range $8.4 < 12 + \log
       \mathrm{O/H} < 8.8$, but at lower metallicity, the
      discrepancy between model and observations can be up to and
      order of magnitude. At higher
       than solar metallicity where an AGN contribution to the
       $\heii{\lambda 4686}$ flux is more likely we see that model also fail to
       predict the same ratio as the observed value.}

\label{fig8}
\end{figure}
\section{Model predictions}
\label{sec:prediction}
In the previous section we carried out an empirical analysis of the
properties of star-forming galaxies in the SDSS which show strong
nebular $\heii{\lambda4686}$ emission in their integrated spectra.
Now we will build on the preceding to explore whether current stellar
models can be used to explain the $\heii{\lambda4686}$ emission seen
in the spectra of these galaxies. We start by predicting nebular
$\heii{\lambda4686}$ emission for galaxies in our sample with the CL01
model. Then we change the stellar population model and explore the
effect of changing the model on the predicted $\heii{\lambda4686}$
line flux.

\begin{figure*}
\centerline{{\includegraphics[width=0.7\textwidth, angle=90]
             {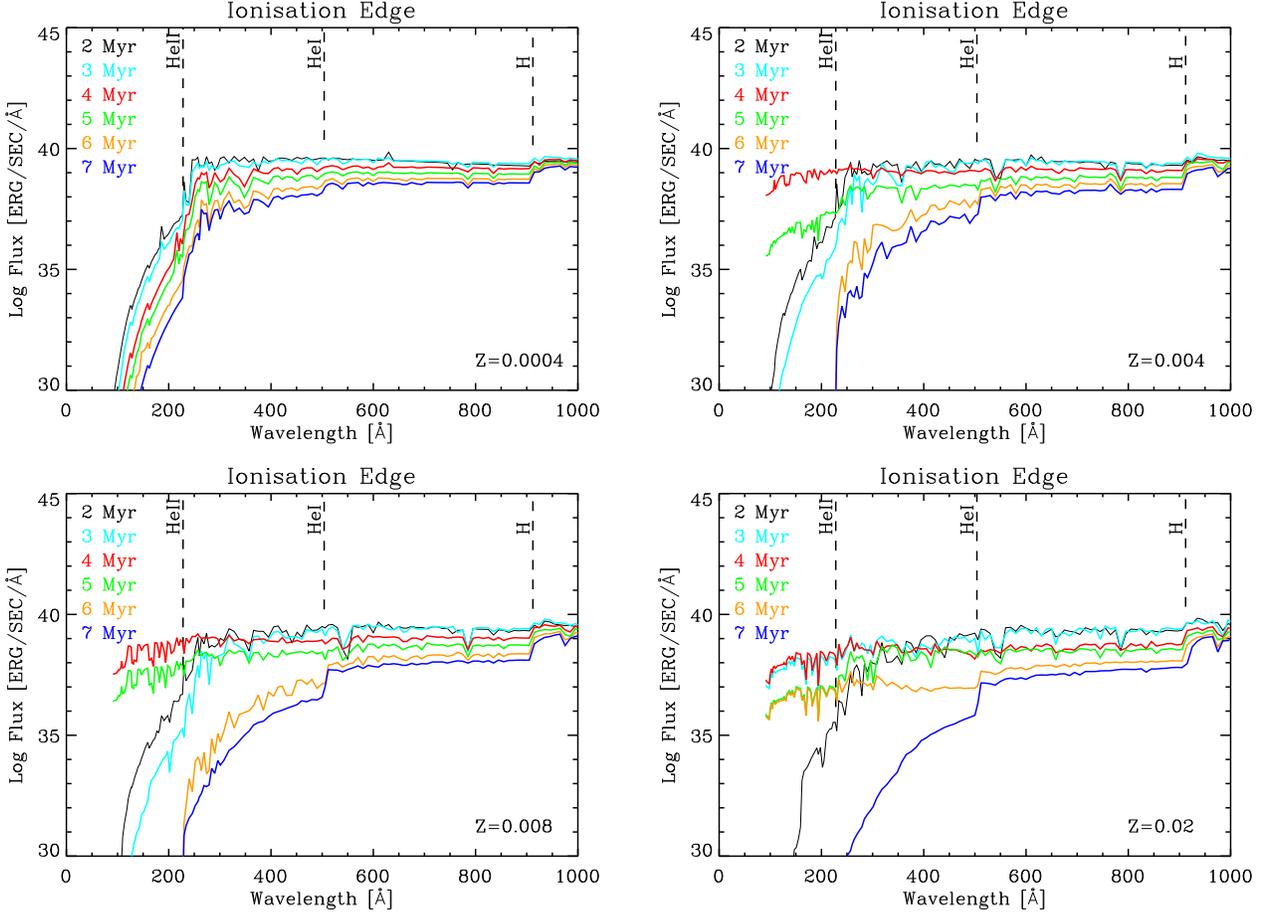}} }
         \caption{This figure shows the spectral energy distribution
           (SED) of an instantaneous burst calculated with Starburst99. 
           Each panel corresponds to one metallicity and shows the
           SED for a range of ages as indicated. Note that the
           appearance of WR stars 4 Myr after the burst results in a
           much harder UV continuum. After 5-6 Myr the WR stars
           disappear and the UV continuum rapidly fades. Also note
           that in these models, the lowest metallicity SED does not
           show a significant WR phase.}
\label{fig9}
\end{figure*}
\subsection{CL01 predictions for nebular He\,II emission}
\label{sec:fitting}

We follow the same procedure as in the calculation of PDFs for the
galaxy parameters in the previous section and calculate the likelihood
of the model for each object in our sample by fitting the CL01 grid of
models to the five important $\rm{\oii{\lambda3727,3729}}$, $\hb$,
$\oiii{\lambda5007}$, $\ha$, $\nii{\lambda6584}$
emission lines.

We now want to see whether the models that reproduce the main strong
lines in the optical spectrum also reproduce the $\heii{\lambda 4686}$
emission line strength. We build the likelihood distribution of
$\heii{\lambda4686}$ flux for each galaxy in the same way as before by
weighting the $\heii{\lambda 4686}$ flux in each model by the
probability of that model. We take the median of the likelihood
distribution as a prediction for nebular $\heii{\lambda4686}$ emission
and the associated confidence interval to be the 16th-84th percentile
range. We follow the same approach to estimate the  $\heii{\lambda
  4686}/\hb$ ratio.

In Figure~\ref{fig8} we show the logarithm of the ratio of this model
prediction for $\heii{\lambda4686}/\hb$ to the observed
$\heii{\lambda4686}/\hb$ ratio as a function of oxygen abundance.  It
is clear that there is acceptable agreement between the model
predictions and the observations in the range $8.4 < 12 + \log
\mathrm{O/H} < 8.8$, but a model that can reproduce most the strong
lines in the spectrum well, predicts up to one order of magnitude
lower $\heii{\lambda4686}/\hb$ ratio than the observed value for some
objects at lower metallicities. We also see a deviation at high
metallicity --- in this regime an AGN contribution to the \heii{} line
flux is more likely, both because the galaxies are more massive, and
also because the star formation-AGN separation is more gradual in this
regime. We will not discuss this mismatch further here.

The population synthesis model used in the CL01 models approximate the
WR emission as black bodies at their effective temperature, note in
passing that this is not the case in the current BC03 models. This
will overestimate the hardness of the ionizing spectra compared to
models that consider more sophisticated WR atmosphere models such as
e.g. Starburst99 \citep{Le} and BPASS \citep{El}.
Given the relatively simple treatment of the WR phases in the CL01
models, one might be concerned that the failure to match the data is
due to an inherent weakness of the models. In the following we
therefore look at the effect of different stellar evolution and
atmosphere models on the prediction of $\heii{\lambda 4686}$
emission line strengths.

\begin{figure*}
\centerline{{\includegraphics[width=0.7\textwidth, angle=90]
              {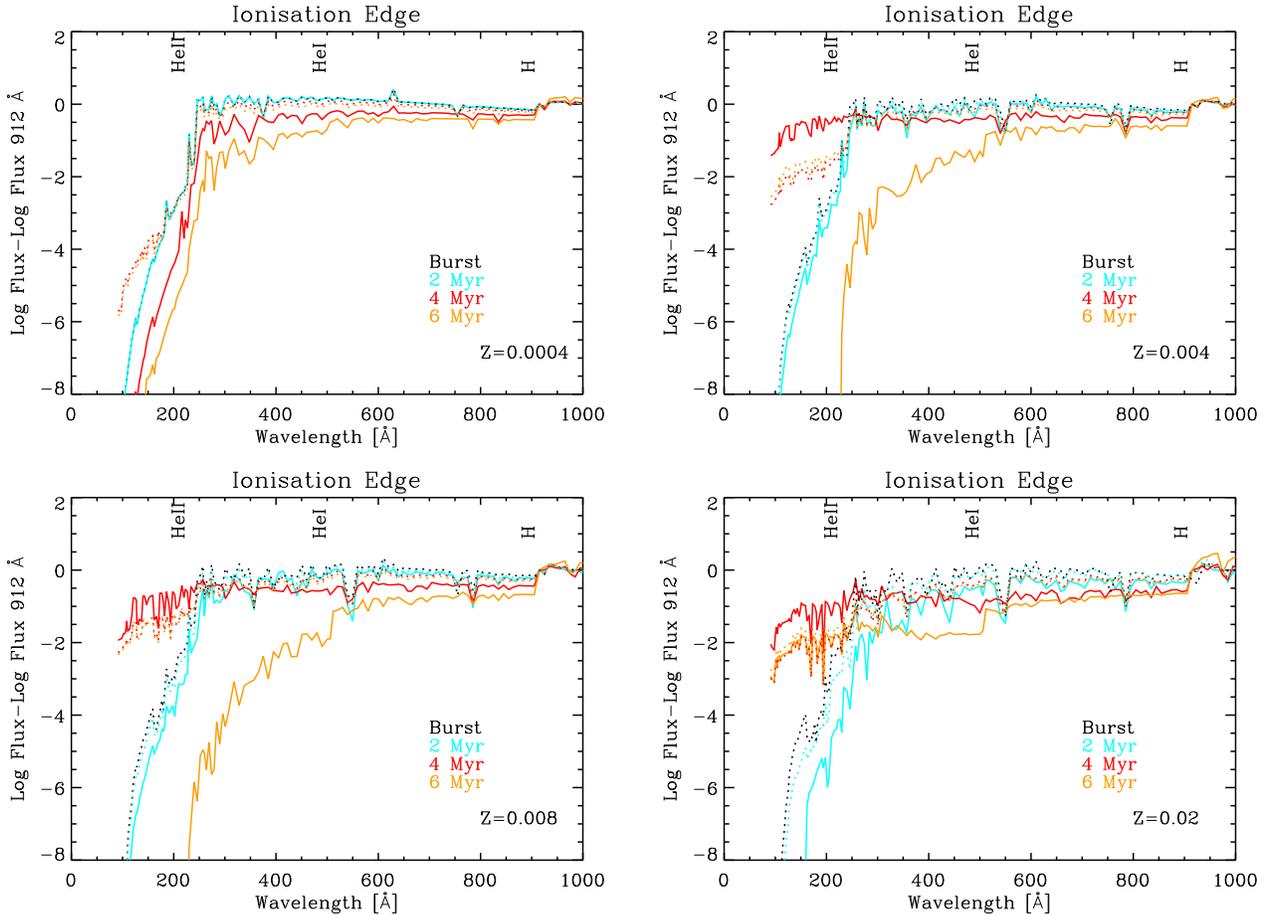}} }
          \caption{Each panel compare the calculated SED of two SFHs with starburst99
            for a different metallicity, solid lines show SEDs of
            instantaneous burst and dotted lines show SEDs with
            constant star formation rate and different colours show
            different burst ages. Fluxes have been normalised to the
            flux at 912 \AA. We can see although models with constant
            star formation form WR stars continuously after 4 Myr, the
            overall shape of the UV continuum is softer than in the
            instantaneous burst models because the continuous
            formation of luminous O stars softens the extreme UV
            spectrum for a fixed rate of hydrogen ionizing photons.}
\label{fig10}
\end{figure*}
%
\subsection{Starburst99 predictions}

To better understand the origin of the $\heii{\lambda{4686}}$ emission
and its dependence on the stellar models adopted, we use the latest
version (6.0) of the spectral synthesis code Starburst99
\citep[Stb99,][]{Le, Le10} to calculate spectral energy distributions
(SEDs) predictions for a range of ages and metal abundances. We
calculate models with an instantaneous burst and a constant star
formation history with a Kroupa IMF.
We have explored a range of stellar evolution models but the
differences are small so we only show the results of one model. For
this we adopt the Padova AGB evolutionary tracks combined with
Pauldrach/Hillier atmospheres \citep{smith}, Stb99 uses O star model
spectra from \citet{Pa01} and WR model spectra from the code of
\citet{Hi98}.  The models include stellar and nebular continuum. We
create models with different metallicities (0.0004, 0.004, 0.008,
0.02, 0.05), where the reference solar metallicity is $Z_{\odot}=0.02$
\citet{An}.  We do not consider dust and run the models up to 100 Myr
with time-steps of $10^5$ years.  Figure~\ref{fig9} shows the
resulting SEDs for an instantaneous burst with a range of
metallicities. Each panel corresponds to one metallicity as indicated
and shows the time of the SED for six burst ages. The plots show
clearly that the appearance of WR stars, 4 Myr after the burst,
results in a harder UV continuum shortwards of the He$^+$ ionizing
edge at 228 \AA. After $\sim 5$\,Myr the WR stars disappear and the UV
continuum becomes softer. This is in good agreement with the
discussion in Schaerer \& Vacca (1998, see their Figure 9), which is
natural as those models lie at core of the WR modeling in Stb99.
Figure \ref{fig10} compares the SEDs of an instantaneous burst (solid
line) with that of a continuous star formation model (dotted line). The SEDs
have been normalised at 912\AA, so have the same amount of hydrogen
ionizing photons. The continuous star formation models form WR stars
continuously after 3 Myr, but the overall shape of the UV continuum is
softer than for instantaneous burst models because of the continuous
formation of luminous O stars which dilute the SED for a given total
mass.

\begin{figure*}
\centerline{\hbox{\includegraphics[width=0.35\textwidth, angle=90]
             {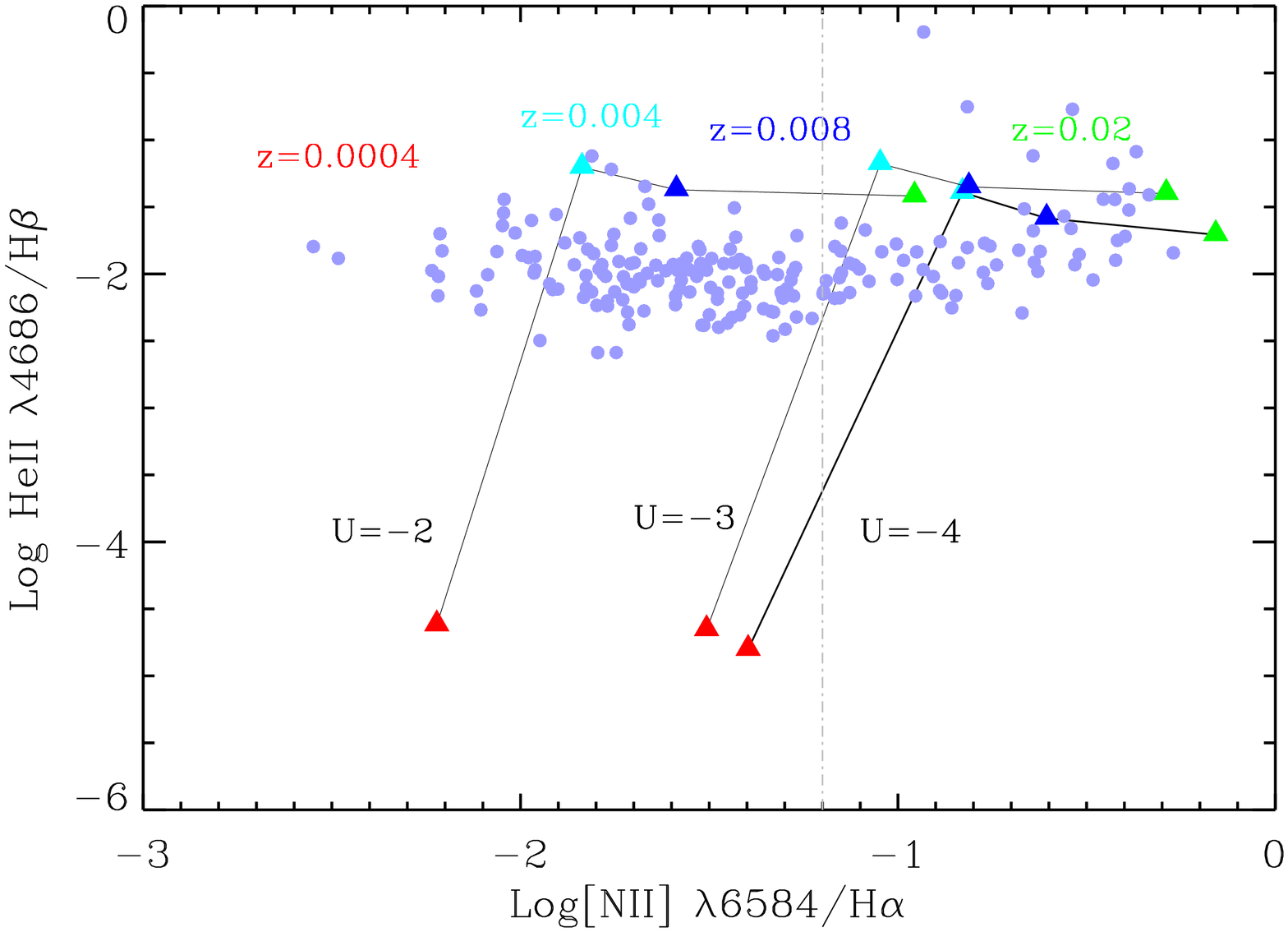}} 
            \hbox{\includegraphics[width=0.35\textwidth, angle=90]
             {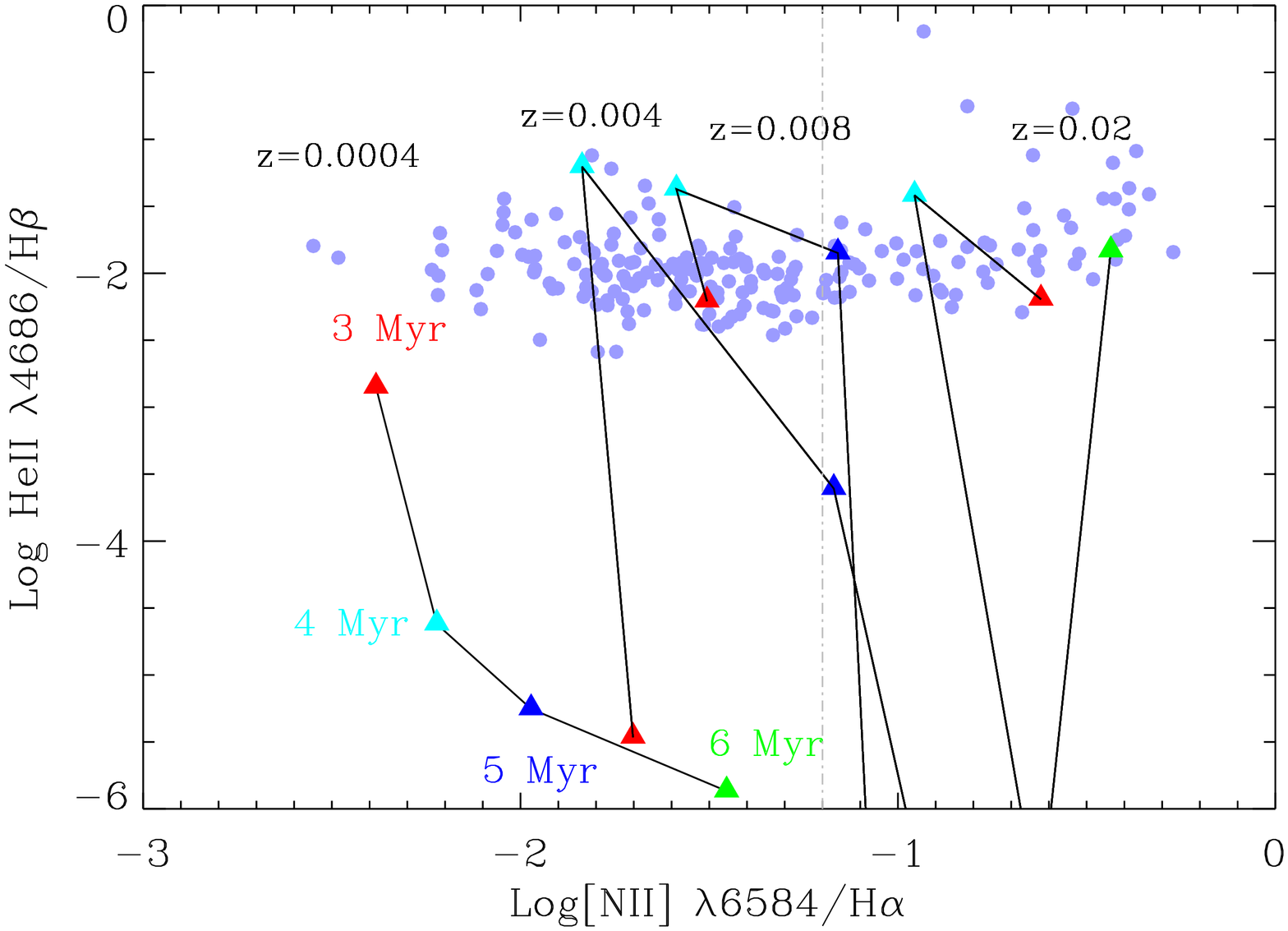}}}
         \caption{The left panel shows the starburst99 instantaneous burst model
           prediction for the $\heii{\lambda4686}/\hb$ ratio for
           different ionization parameters at a fixed age of 4 Myr.
           The triangles along each line correspond to the
           metallicities (0.02, 0.2, 0.4, 1) $Z_{\odot}$ and our
           sample galaxies are shown as purple circles.  The right
           hand panel shows the same for a range of different
           metallicities and ages with $\log \mathrm{U}$ set to
           $-2$. The triangles along each model line correspond to the
           burst ages (3, 4, 5, 6) Myr. Note that the lowest
           metallicity model is unable to cover the observational
           data. The grey dashed-dotted lines in each panel show the
           N2 PP04 metallicity calibration for $\rm{12+\log O/H}=8.2$}
\label{fig11}
\end{figure*}

To calculate the emission lines, we use the UV continuum generated by
the Stb99 models as an input to the photoionization code Cloudy
\citep[version c08,][]{Fer}. For each time step, ionization bounded
models are calculated by varying the ionization parameter
$\mathrm{\log U} = $ $-2$, $-3$, $-4$, where for consistency with the
CL01 models we calculate the ionization parameter at the edge of the
Str{\"o}mgren sphere, and a constant hydrogen density of $\log
\rm{n_{H}/cm^{-3}} = 2.5$. This range in ionization parameter spans
the range found by \citet{Stas96} in their analysis of intensely
star-forming galaxies (approximately $-3.5<\log \mathrm{U} < -2.5$).
Figure~\ref{fig11} shows the calculated $\heii{\lambda4686}/\hb$ ratio
versus $\nii{\lambda6584}/\ha$ for different metallicities and
different ionization parameters in the left panel and different burst
ages in the right panel.  The triangles along each model line
correspond to the metallicities (0.02, 0.2, 0.4, 1) $Z_{\odot}$ in the
left panel and to ages of 3, 4, 5, and 6 Myr in the right panel. In
the left panel we fix the the age to 4 Myr, and in the right we set
$\log \mathrm{U}=-2$.  The galaxies in our sample are shown as filled
purple circles.  From these plots we can see how the
$\heii{\lambda4686}/\hb$ ratio depends on age, metallicity and
ionization parameter.  The lowest metallicity considered in this work
is $Z=0.02$ $Z_{\odot}$ and for this metallicity there is a strong
discrepancy between the model predictions and the observed data but
predictions for other metallicities at 4 Myr agree well with the
observed $\heii{\lambda4686}/\hb$ ratios.
The models can predict the $\heii{\lambda4686}$ emission line ratio,
but only for instantaneous bursts with metallicity of 20\% solar and
above, and only for ages of $\sim 4-5$ Myr, the period when the
extreme-ultraviolet continuum is dominated by emission from WR
stars. For burst ages younger than 4 Myr and older than 6 Myr, and for
models with a continuous star formation (not shown here), the
softer ionizing continuum results in an emission spectrum that has too
weak \heii{} lines to be consistent with the observational data.
\begin{figure*}
\centerline{\hbox{\includegraphics[width=0.7\textwidth, angle=90]
             {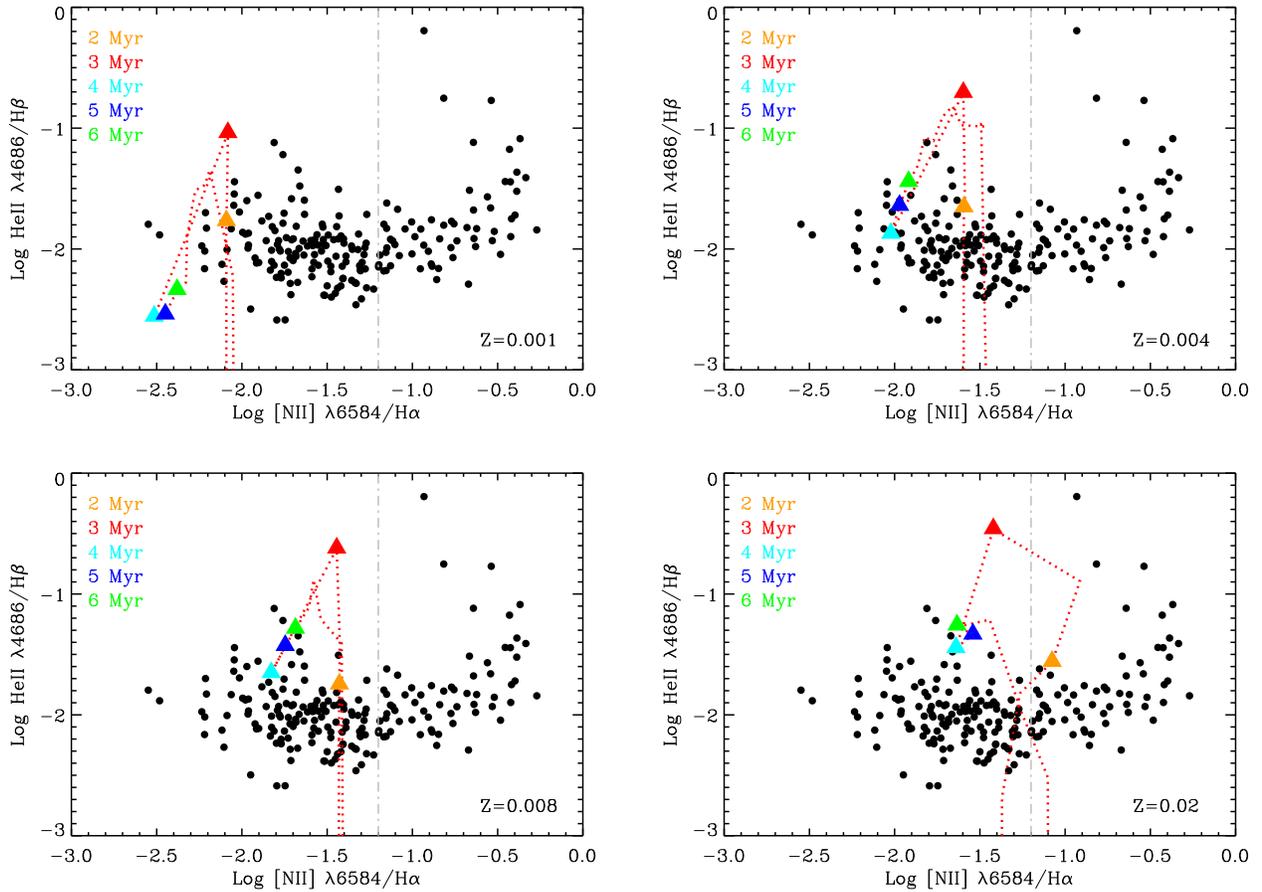}} }
\caption{The plots show the prediction of instantaneous burst binary model (BPASS code) 
for $\heii{\lambda4686}/\hb$ ratio for different metallicities 
(0.05, 0.2, 0.4,1) $z_{\odot}$ and burst ages. The grey dashed-dotted line shows 
the N2 PP04 metallicity calibration for $\rm{12+\log O/H}=8.2$.} 
\label{fig16}
\end{figure*}
\begin{figure*}
\centerline{\hbox{\includegraphics[width=0.7\textwidth, angle=90]
             {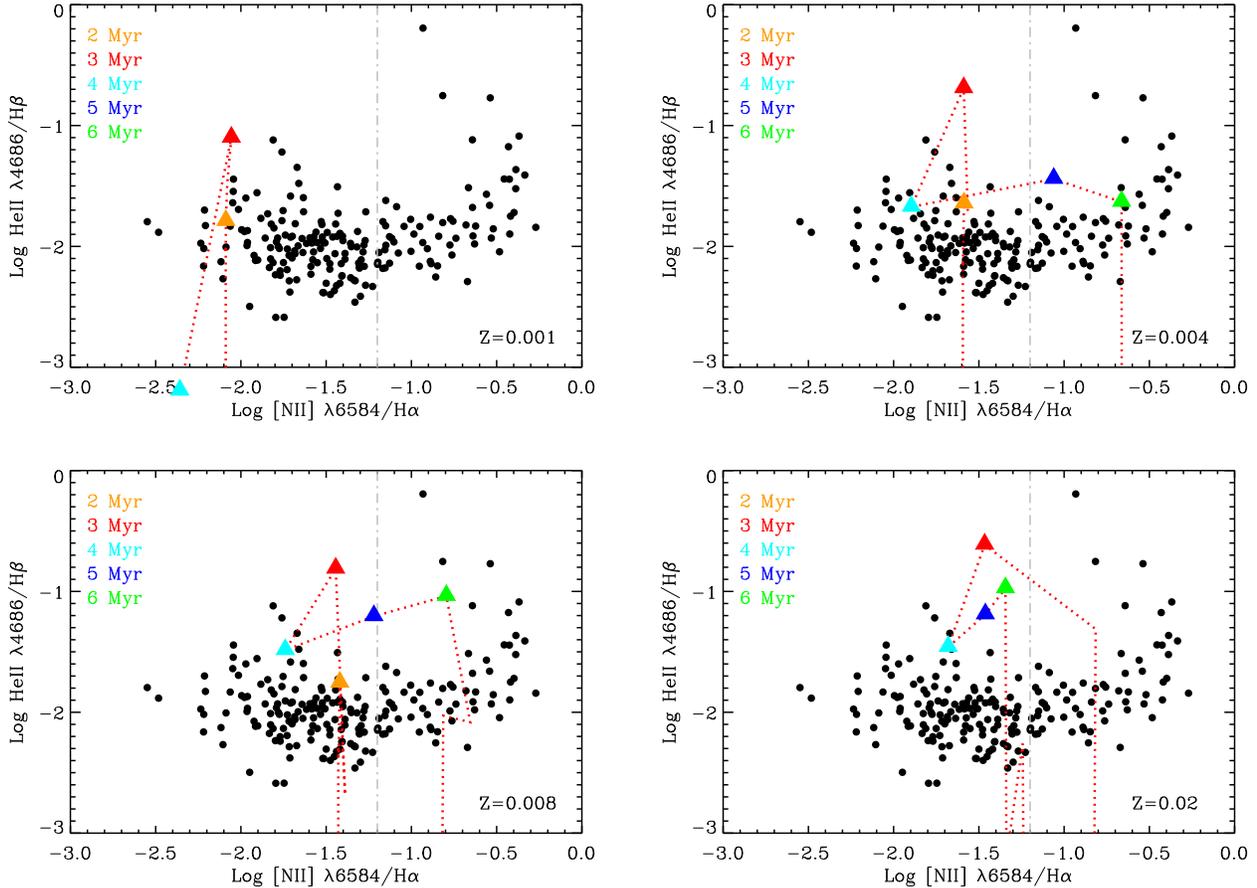}} }
         \caption{The plots show the prediction of instantaneous burst
           single-star model using the BPASS code for the
           $\heii{\lambda4686}/\hb$ ratio for different metallicities,
           (0.05, 0.2, 0.4, 1) $Z_{\odot}$, and burst ages. The grey
           dashed-dotted line shows the N2 PP04 metallicity
           calibration for $\rm{12+\log O/H}=8.2$.}
\label{fig17}
\end{figure*}
\subsection{The effect of binary evolution on the He II 4686 emission}

The Stb99 models consider single-star evolution only, but it is
well-known that massive stars are frequently found in binaries and
higher order systems which can have a major effect on the evolution of
massive stars.  To explore this possibility we compare the observed
$\heii{\lambda4686}/\hb$ ratio to the prediction of the Binary
Population and Spectral Synthesis (BPASS, Eldridge et al 2008, 2009,
2011)\nocite{El,El09,El11} model. The BPASS code includes a careful
treatment of the effect of binary evolution on massive short lived
stars, and Eldridge et al found that including massive binary
evolution in the stellar population leads to WR stars forming over a
wider range of ages up to 10 Myr which increases the UV flux at later
times.  In Figure \ref{fig16} and Figure \ref{fig17} we show the observed
$\heii{\lambda4686}/\hb$ ratio in comparison with their instantaneous
burst binary and single-star population models, respectively.  Each
panel shows four different metallicities (0.05, 0.2, 0.4,1)
$Z_{\odot}$. The grey dashed-dotted line shows the N2 PP04 metallicity
calibration for $\rm{12+\log O/H}=8.2$. We should note that the lowest
metallicity in these plots is a factor of 2.5 higher than that of
Stb99 (0.02 $Z_{\odot}$).

We get the highest value for the $\heii{\lambda4686}/\hb$ ratio at 3
Myr and the period with elevated $\heii{\lambda4686}/\hb$ lasts
longer.  Comparing Figure \ref{fig16} and Figure \ref{fig17}, there is not a
striking difference in the predicted peak ratio for
$\heii{\lambda4686}/\hb$ between single-star and binary population
models but clearly the binary model predict an elevated ratio for a
longer period of time than the single-star models.
\begin{figure*}
\centerline{\hbox{\includegraphics[width=0.7\textwidth, angle=90]
             {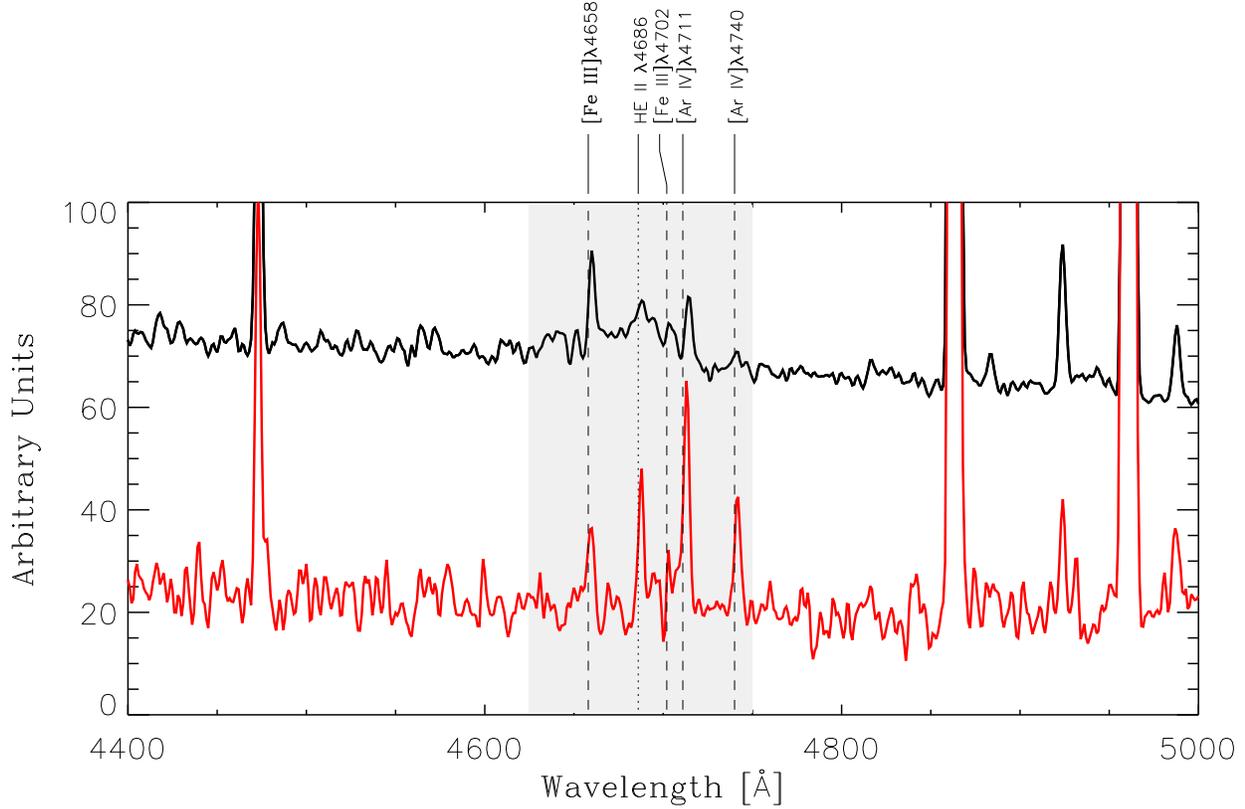}}} 
         \caption{This plot shows a comparison between two stacked
           spectra of galaxies with \heii{} emission. The black
           spectrum shows the result of stacking the spectra of all
           galaxies in our sample that show WR features. In this case
           we can clearly see the blue WR bump (gray shaded
           region). The red spectrum shows the result of stacking the
           spectra of all galaxies that show no WR features. Despite
           the increase in S/N we see no sign of WR features,
           strengthening our claim that this class of galaxies show no
           signs of WR features.}
\label{fig22}
\end{figure*}
\section{The origin of nebular He II 4686 emission}
\label{sec:origin}
The preceding discussion makes it clear that in standard models for
stellar evolution, only during the phases where the ionizing spectrum
is dominated by WR stars will we see strong $\heii{\lambda4686}$
emission. This was already pointed out by \cite{Sch96} and our results
are in good agreement with that work as well as a number of other
previous works \citep{Sch98,Gu00,Thu05}.  

What has been less studied is the direct test of this prediction,
namely to ask whether WR features are seen whenever
$\heii{\lambda4686}$ is observed.  \cite{Keh11} and \cite{NM} have
recently studied individual star forming regions with
$\heii{\lambda4686}$ emission in the Local Group and have shown that
while most \heii-emitting regions also show evidence of WR stars,
not all do. In more distant galaxies, previous efforts have primarily
looked at \heii{} emission in samples selected for other purposes. As
an example, B08 looked at \heii{} galaxies in SDSS DR6 and studied
whether they showed WR features, but their sample selection was not
optimised for finding galaxies with $\heii{\lambda4686}$ emission. The
excellent analysis of \citet{Thu05} is another example, where the
focus was on BCD galaxies.  Since the present sample is selected
purely on the presence of $\heii{\lambda4686}$, we can carry out this
study with more reliability but we need to check for WR features in
the spectra of our galaxies.

The presence of WR stars can be recognised via the WR bumps around
$\lambda4650$ \AA (blue bump) and $\lambda5808$ \AA (red bump). The
former is a blend of $\heii{\lambda4686}$ and several metal lines
while the latter is caused by $\mathrm{C\textsc{iv}}\lambda5801-12$
(see \citet{crow07}, B08 and references therein).  We have therefore inspected the
spectra in our star-forming sample following the methodology described
in B08 (see Table~\ref{table1}). Each spectrum is assigned a
classification of 0 (no WR), 1 (possible WR), 2 (very probable WR), or
3 (certain WR).  We will consider any spectrum with class 1, 2 or 3 to
show WR features. To check the reliability of these classifications we
can check whether duplicate observations of the same object are given
the same classification. There are four sets of duplicate observations
(0417-51821-513 Class 2, 0418-51817-302 Class 3, 0418-51884-319 Class
2), (0455-51909-073 Class 0, 0456-51910-306 Class 0), (0266-51602-089
Class 1, 0266-51630-100 Class 0) and (0308-51662-081 Class 3,
0920-52411-575 Class 3). Only for 0266-51602-089 and 0266-51630-100 is
there uncertainty whether spectrum show evidence of WR features or not
--- we choose to keep the original classifications. This is in
agreement with, but somewhat better than what we find for the full
sample of WR galaxies from the B08 sample where the RMS classification
uncertainty from duplicates is 0.4 classes without any apparent
dependence on the median S/N of the spectra down to S/N$\sim 10$; at
lower S/N we do not have sufficient numbers of duplicate observations
to make a statement. In total we find that 116 of objects show WR
features.

To check whether the non-detection of WR features is due to a low
S/N, we have created stacked spectra of galaxies with and without WR
features. The result of this exercise is shown in in Figure
\ref{fig22}. The black line is the stack of spectra that show WR
features while in red we show the stack of non-WR galaxies. We see no
sign of WR features in the stack, further supporting the notion that
this class of objects show no signs of WR features.

\begin{figure}
\centerline{\includegraphics[width=0.35\textwidth, angle=90]
             {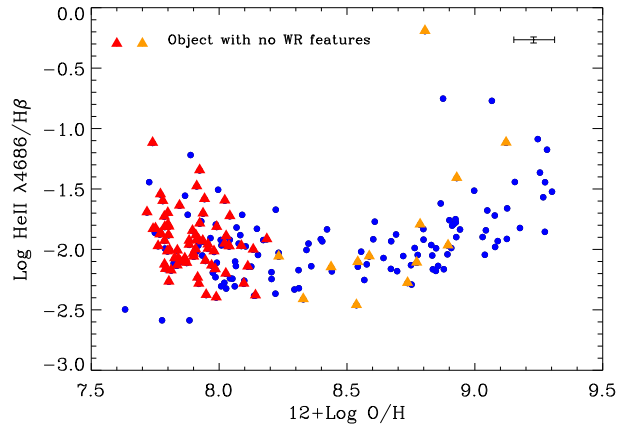}}
           \caption{The $\heii{\lambda4686}/\hb$ ratio versus oxygen
             abundance for the sample. The blue solid circles show the
             location of galaxies with WR features. Galaxies which do
             not show WR features are indicated by red and orange
             triangles at lower and higher oxygen abundances than
             $12+\log \rm{O/H}=8.2$, respectively.}
\label{fig23}
\end{figure}

We now turn to explore whether there are physical differences in the
galaxies showing WR features or not. Figure \ref{fig23} shows the
ratio of $\heii{\lambda 4686}/\hb$ versus oxygen abundance for
galaxies with (blue circles) and without WR features (triangles). The
symbols for the non-WR galaxies, here and in the following, are
coloured red for $12+\log\mathrm{O/H}< 8.2$, and orange otherwise.  At
oxygen abundances lower than 8.2 we see there is a trend of increasing
$\heii{\lambda4686}/\hb$ ratio towards lower metallicities. We
interpret this as being due to a harder ionizing continuum at lower
metallicities \citep[e.g.][ and references therein]{Thu05}. We also
see a trend of increasing $\heii{\lambda4686}/\hb$ towards higher
metallicity for $12 + \log \mathrm{O/H} > 8.7$.  It is less clear what
causes this, but since these are more massive systems with higher star
formation rates and with stronger stellar winds due to their higher
metallicity, it is likely that what we are seeing is due to an
increased contribution of shocks and/or a low-level AGN contamination.

Another striking result is that for $12 + \log\mathrm{O/H} > 8.2$,
essentially all \heii-emitting galaxies show WR features, in
agreement with what one would expect from the models discussed in the
previous section. At high metallicity there appear to be some systems
that show a high $\heii{\lambda4686}/\hb$ ratios but no sign of WR
stars. Since these are more massive systems, and fall intermediate
between the SF and AGN groups in the $\heii{\lambda4686}/\hb$ versus
$\nii{\lambda6584}/\ha$ diagram, we interpret this as a likely sign of
a low-level AGN contribution.

\begin{figure}
\centerline{{\includegraphics[width=0.35\textwidth, angle=90]
    {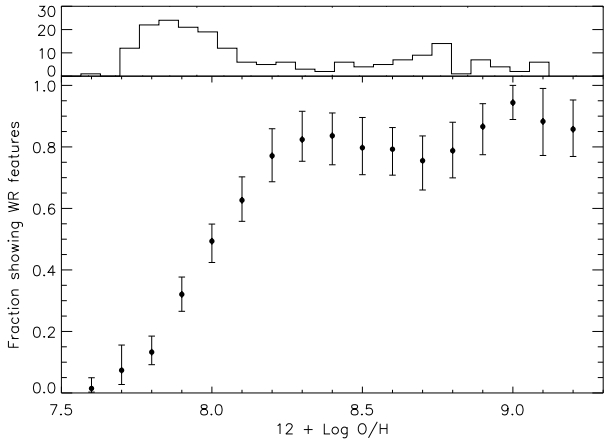}} }
\caption{The fraction of objects with detected WR features in the \heii\
  sample as a function of gas-phase oxygen abundance. The points
  show the median fraction in each abundance bin and the error bars
  the 16\%-84\% scatter around the median (see text for
  details). While essentially all high-metallicity star-forming galaxies with
  $\heii{\lambda 4686}$ nebular emission show WR features, this
  fraction drops rapidly at metallicities below $12+\log \rm{O/H}
  \approx 8.2$.}
\label{fig13}
\end{figure}

In Figure~\ref{fig13} we show the fraction of galaxies that show WR
features as a function of metallicity. To calculate this we draw a
number of random realisations of the data using the uncertainty
estimates on the oxygen abundance to draw a random realisation. We
also draw a random realisation of the WR classification assuming a
random uncertainty in the classification of 0.4 as determined from the
analysis of duplicate spectra as discussed above. We carry out 101
random realisations for each of the 101 bootstrap repetitions and
calculate the median fraction of galaxies showing WR features in
each metallicity bin as well as the 16\%--84\% scatter around the
median which is shown by the error bars in Figure~\ref{fig13}. What is
clear is that there is a transition at around an oxygen abundance of
$12 + \log \mathrm{O/H}\approx 8.2 \pm 0.1$. The uncertainty of $0.1$
dex encapsulates the fact that the exact value of this transition
abundance depends somewhat on the metallicity calibration adopted and
0.1 dex corresponds to the scatter found when using the different
calibrations discussed earlier.

\begin{figure}

\centerline{\hbox{\includegraphics[width=0.35\textwidth, angle=90]
             {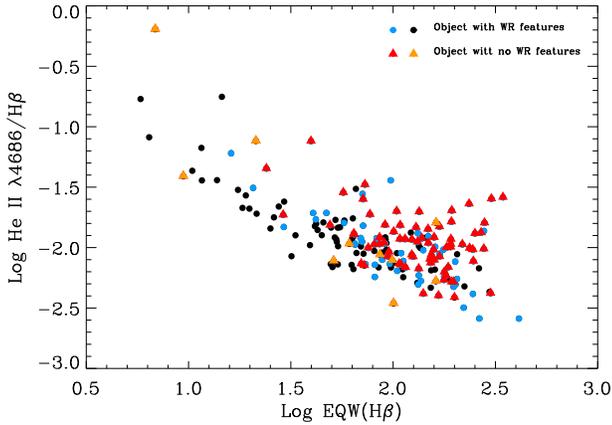}}}  
         \caption{This plot shows the dependence of the
           $\heii{\lambda4686}/\hb$ ratio on the $\mathrm{EW(\hb)}$
           and metallicity. We show galaxies with WR features by black
           and blue circles at high and low metallicities,
           respectively. Galaxies without WR feature are shown by
           orange and red triangles at high and low metallicities.  We
           see the $\heii{\lambda4686}/\hb$ ratios decrease with
           $\mathrm{EW(\hb)}$ for both WR and non-WR objects.
           However, galaxies without WR features show higher ratios
           than galaxies with WR feature for the same
           $\mathrm{EW(\hb)}$, especially at lower
           $\mathrm{EW(\hb)}$s, alternatively one could read this to
           say that they have a high $\mathrm{EW(\hb)}$ (young age)
           for a given $\heii{\lambda4686}/\hb$ ratio.}
 \label{fig14}
\end{figure}

As we showed in earlier, e.g.\ Figure~\ref{fig11}, the
$\heii{\lambda4686}/\hb$ ratio depends strongly on the age of the starbursts
so it is reasonable to ask whether the systems without WR features are
systematically younger or older than the systems that show WR
features.  In Figure~\ref{fig14} we test this by plotting the
$\heii{\lambda4686}/\hb$ ratio versus $\mathrm{EW(\hb)}$ for the
sample, where we take the $\mathrm{EW(\hb)}$ as a proxy for starburst
age. We show galaxies with WR features by black and blue circles at
high and low metallicities, respectively. Galaxies without WR features
are as before shown by orange and red triangles at high and low
metallicities.  We see the $\heii{\lambda4686}/\hb$ ratios decrease
with $\mathrm{EW(\hb)}$ for both WR and non-WR objects. However,
galaxies without WR features show higher ratios than galaxies with WR
feature for the same $\mathrm{EW(\hb)}$, especially at lower
$\mathrm{EW(\hb)}$. Alternatively one might say that at a fixed
$\heii{\lambda4686}/\hb$ the systems without WR features have a higher
$\mathrm{EW(\hb)}$, or with our assumption, a younger starburst. It is
not possible with our data to disentangle these two possibilities.

\section{Why are there galaxies with He\,II 4686 emission but no WR features?}
\label{sec:explanation}
The models that we discussed previously agree that WR stars are the
source of the hard ionizing photons necessary to produce the
$\heii{\lambda4686}$ emission. It is therefore a puzzle why some of
the galaxies with nebular $\heii{\lambda4686}$ emission do not show
stellar WR features. This lack of WR features has been pointed out
before \citep[e.g][]{Thu05}, but this is the first time a clear trend
with metallicity appears.  In this section we therefore turn to
discuss some possible reasons for this lack of WR features.

\begin{figure}
\centerline{\hbox{\includegraphics[width=0.35\textwidth, angle=90]
    {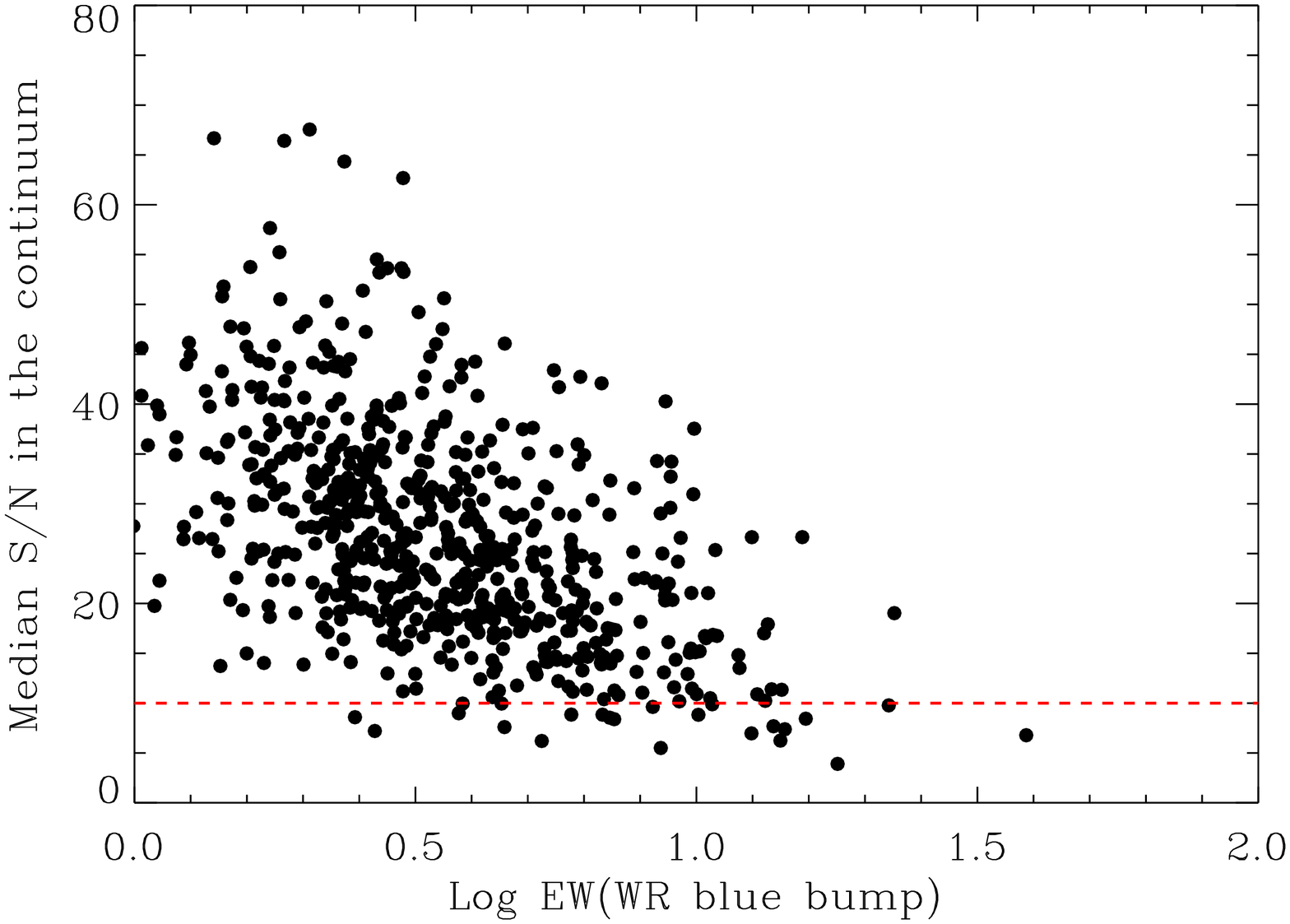}}}
\centerline{\hbox{\includegraphics[width=0.35\textwidth, angle=90]
    {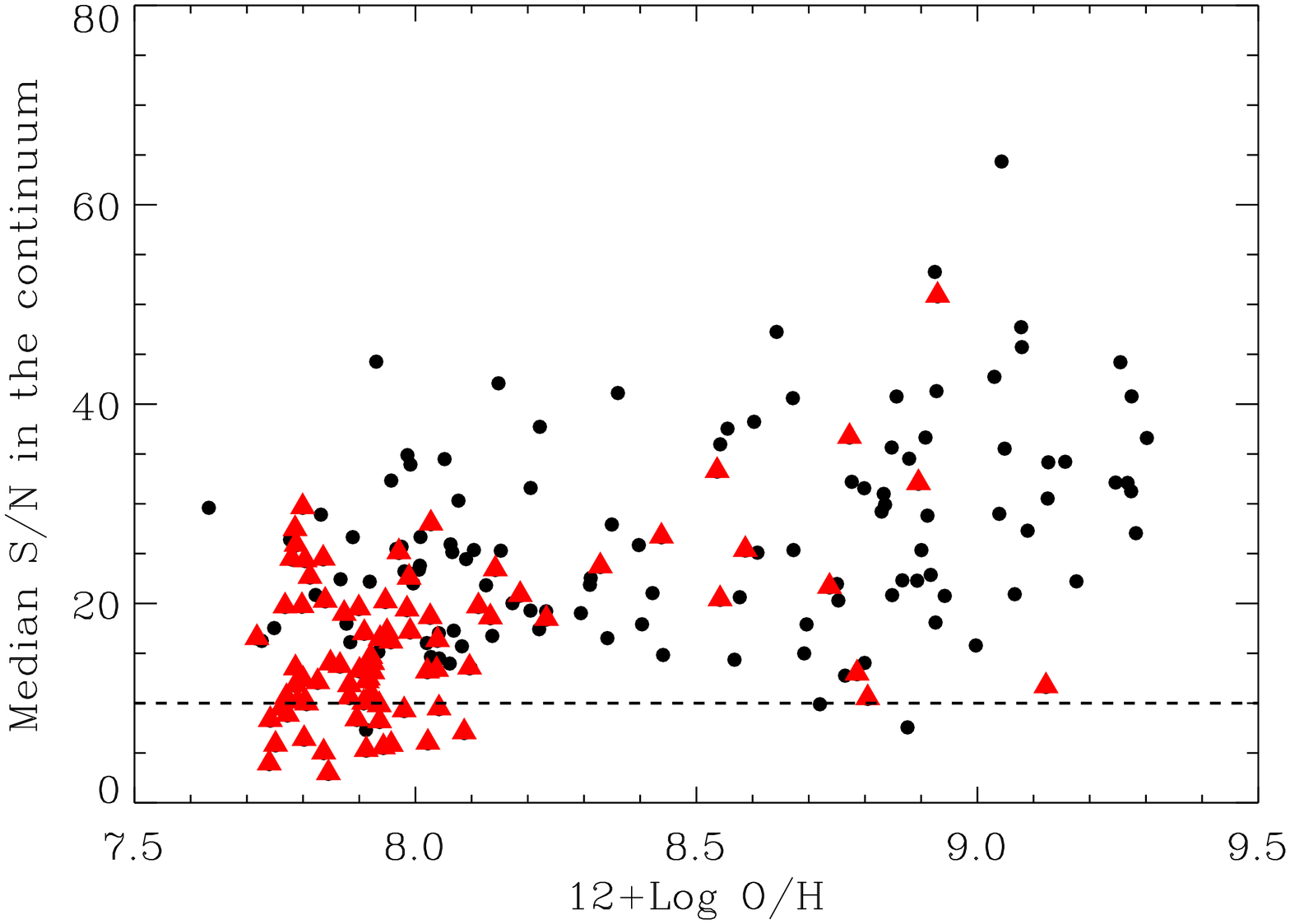}}}
\caption {Top panel: The relationship between the equivalent width of the WR blue
  bump and the median S/N in the continuum for WR galaxies in SDSS
  DR7. We see that a S/N of 10 is sufficient to detect 
  even very weak features. Bottom panel: The distribution of the
  median S/N in the continuum for WR and non-WR 
  objects in the sample versus their oxygen abundance, we see that 20
  out of 70 of the non-WR galaxies at $12 + \log\mathrm{O/H} < 8.2$
  have S/N less than 10. See text for more discussion.}
\label{fig15}
\end{figure}

\noindent \textbf{1. Differences in the $\mathbf{S/N}$ of the spectra}\\
To detect the WR features we need fairly high S/N in the continuum as
they are broad and weak features. In the top panel in
Figure~\ref{fig15} we show the relationship between the equivalent
width of the WR blue bump and the median S/N in the continuum for WR
galaxies in SDSS DR7. We see that a S/N of 10 is sufficient to detect
even very weak features. The distribution of the S/N for WR and non-WR
objects in the sample versus their metallicities in the bottom panel
shows that 20 out of 70 of the non-WR galaxies at $12 +
\log\mathrm{O/H} < 8.2$ have S/N less than 10. Since the total number
of galaxies at $12 + \log\mathrm{O/H} < 8.2$ is 115, if all the 20 low
S/N galaxies are assumed to have WR features this brings the fraction
of galaxies with WR features in this metallicity range from 30\% to
43\%. Thus low S/N could cause us to underestimate the WR fraction by
at most $\sim 15$\%.  But the fact that the co-added spectrum in
Figure~\ref{fig22} which has a higher S/N shows no WR features is
suggestive that S/N might not be the problem for non-detection of WR
features.

Furthermore even adding 15\% would still mean that less than half
\heii{} emitters at low metallicity would show WR features.  We also
studied the difference in WR classification of duplicate observations
as a function S/N and as remarked earlier, we saw no trend with S/N
for more uncertain classification down to a S/N of 10. The dataset is
insufficient to test this at lower S/N. Thus our conclusion is that it is
unlikely that the systematic absence of WR features in low metallicity
objects is due to low S/N in the spectra.

\noindent \textbf{2. Weak lined WN stars}\\
One possible reason for the lack of WR features in the most metal poor
galaxies, is that the WR lines are too weak to be seen. It is well
known \citep[e.g.][]{conti,crow06} that WR stars in the SMC have
narrower and less luminous lines than equivalent stars in the Milky
Way. \citet{crow06}, for instance, find that He II lines in WR stars
in the SMC are typically a factor of 4-5 weaker than the Milky Way and
hence one would expect that in galaxies at the same distance, it would
be harder to see WR features in lower metallicity systems. This is
however countered by the fact that low metallicity systems on average
are closer, and hence the SDSS fibre subtends a smaller physical
size. Since WR emitting regions typically are small, they do not fill
the 3" aperture in more distant galaxies, which means that the
contrast of the WR features is being enhanced in low redshift
systems. This is reflected in the fact that in the WR survey by B08,
the equivalent width of the WR features in low metallicity systems (12
+ Log O/H $< 8.25$) is slightly higher than that in metal rich objects
(12 + Log O/H $>8.5$), a mean of 5.1 \AA\ vs 4.1 \AA. Thus the
increased contrast appears to approximately cancel out the decrease in
line luminosity leading to a fairly constant detection potential with
redshift. Thus we do not believe that this is the cause of the dearth
of WR features in the low metallicity \heii{} emitting galaxies, at
least down to $12 + \log \mathrm{O/H} \sim 8$. At very low
metallicities, $12 + \log \mathrm{O/H} \sim 7.5$, Eldridge \& Stanway
(2009) found that in their models the WR features become very weak and
if those results are correct it should be extremely hard to detect WR
features in those galaxies. We do however caution that I Zw 18 has
very prominent WR features \citep[e.g.][]{Leg97}, so at least some
extremely metal poor galaxies do show clear WR features. Furthermore,
even if we are unable to detect WR features at the very lowest
metallicities, the same models predict strong WR features at $12 +
\log \mathrm{O/H} > 8$, thus this is not a sufficient explanation for
the result in Figure~\ref{fig13}.
\\
\noindent \textbf{3. Shocks}\\
\citet{Thu05} studied the hard ionizing radiation in very metal poor
BCD galaxies in the local Universe and concluded that fast radiative
shocks could be responsible for the nebular $\heii{\lambda4686}$
emission. Therefore, another possibility is that there is a
contribution to the $\heii{\lambda4686}$ from shocks in the ISM
\citep{Dop96}.  That $\heii{\lambda4686}$ can have some contribution
from shocks is plausible, but whether it can explain the systematic
lack of WR features at low metallicity is less clear.

One question is whether shock models can reproduce the line luminosity
in \heii{} at low metallicities. Using the predictions of the
\citet{dop05} and \citet{Allen} shock models for the
$\heii{\lambda4686}/\hb$ ratio versus $\nii{\lambda6584}/\ha$, we
found that we can only obtain a ratio comparable to the observed one
for objects at high metallicity. A second issue centers on the
observation that shocks would most likely come from supernovae and
stellar winds, but the latter are though to be weaker at low
metallicities. Shocks can also be induced by outflows from starbursts
(e.g. galactic winds) and mergers but only
two galaxies in our sample are interacting and show a perturbed
morphology and we see no clear difference between the low- and
high-metallicity subsamples. \\
\noindent\textbf{4. X-ray binaries}\\
Another candidate source for \heii{} ionization that has been
discussed in previous studies of $\heii{\lambda4686}$ emitting nebulae
are massive X-ray binaries \citep{Gar}. While these are likely present
in many active star forming regions, the question here is why massive
X-ray binaries should be more common at low metallicity.  If the
He$^+$ ionizing photons at low metallicities come from X-ray binaries,
we would expect an increasing binary fraction with decreasing
metallicity. Without a theoretical justification for this, we consider
an increased abundance of X-ray binaries at low metallicity to be an
unlikely explanation for the trend seen in Figure~\ref{fig13}. \\
\noindent \textbf{5. post-AGB stars}\\
\citet{Bin94} demonstrated that photoionization by post-AGB stars can
produce nebular $\heii{\lambda4686}$ emission. After about a few
$10^7$ years (i.e, after massive stars disappear), the ionizing
radiation comes from post-AGB stars. So, the ionization from post-AGB
stars become more important in more evolved systems and this is not
the case for our objects especially not for objects at low
metallicities (c.f.\ Figure~\ref{fig14}). \\
\noindent \textbf{6. Spatial offset}\\
One possible explanation for non detection of WR features is that
there could be a significant spatial separation between the WR stars
and the region emitting \heii. \citet{Keh08} saw indeed such a spatial
separation based on integral field spectra of II Zw 70. The
found that the location of the WR stars and the $\heii{\lambda4686}$
emission appear to be separated by
$\sim 80$pc.\\
Similarly, \cite{Izb06} studied two-dimensional spectra of an
extremely metal-deficient BCD galaxy SBS 0335-052E and showed that the
$\heii{\lambda4686}$ emission line was also offset from the near
evolved star clusters but in their case, by studying the kinematical
properties of the ionized gas from the different emission lines they
suggested that the hard ionizing radiation responsible for the
$\heii{\lambda4686}$ emission was not related to the most massive
youngest stars, but rather was related to fast radiative shocks.

If this offset between the WR stars and the region emitting \heii\
might be an explanation for our non-detections of WR features, it
would mean that such a spatial separation is much more common at low
metallicity which is rather surprising, since stellar winds are
thought to be considerably weaker there, however low metallicity
galaxies are also on average closer so the SDSS fibre subtends a
smaller physical scale so a smaller volume would need to be blown out
by the wind (see Figure 18).  To get a more quantitative estimate we calculate the
gravitational binding energy of a cloud with radius 1.5'' (SDSS
aperture radius) at the redshift of each non-WR object.  We assume a
hydrogen density of $\sim$ 50 $cm^{-3}$. The median energy require to
excavate a hole of this size, is of the order of $10^{55}$ erg for our
full sample.  For the lowest redshift objects the energy requirement
is a much more manageable $10^{49}$ erg and thus in these cases the
absence of WR features in the spectrum could be due to a spatial
offset from the \heii{} emitting region. At higher redshift the
energetics makes this a much less likely explanation. The possibility
does however warrant further examination and to test this we are
undertaking
a spectroscopic follow-up of a subsample of these sources.\\
\noindent \textbf{7. Chemically homogeneous stellar evolution}\\
A final explanation could be that the stellar populations at very low
metallicities can have much higher temperatures than is currently
expected in models. This would be the case if some stars rotated fast
enough to evolve homogeneously \citep{Ma87,Me07,Yoon,Yoon06,cant07}.
In that case we can get a higher continuum at 228 \AA\ and
correspondingly a higher $\heii{\lambda{4686}}/\hb$ ratio in
comparison with non-homogeneous stellar evolution models. An appealing
aspect of this, speculative, explanation is that homogeneous evolution
is predicted to be more common at low metallicity. There are however
currently no studies of the nebular $\heii{\lambda 4686}$ line in the
literature.  \citet{El11b} looked at the effect of quasi-homogeneous
evolution in their binary models and showed that including it led to
\emph{strengthened} WR features at low redshift, in contrast to what
we need. However that still leaves the possibility open that there is
a period where strong nebular $\heii{\lambda 4686}$ is seen but no WR features as
that has not yet been tested.

\section{CONCLUSION}
\label{sec:conclusion}
We have presented a sample of rare star-forming galaxies with strong
nebular $\heii{\lambda4686}$ emission spanning a wide range in
metallicity.  We have derived physical parameters for these galaxies
and showed that emission line models that can reproduce the strong
lines in the galaxy spectra are not able to predict the observed ratio
of $\heii{\lambda4686}/\hb$ at low metallicities. In agreement with
previous studies we found that current models for single massive stars
are able to reproduce the $\heii{\lambda4686}/\hb$ ratio in galaxies
in our sample, but only for instantaneous bursts of 20\% solar
metallicity or higher, and only for ages of $\sim 4-5$ Myr, the period
when the extreme-ultraviolet continuum is dominated by emission from
WR stars. For stars younger than 4 Myr or older than 5 Myr, and for
models with a constant star-formation rate, the softer ionizing
continuum results in $\heii{\lambda4686}/\hb$ ratios typically too low
to explain our data. Including massive binary evolution in the stellar
population analysis leads to WR stars occurring over a wider range in
age which leads to acceptable agreement with the data at all
metallicities sampled as long as WR stars are present.

However, the most notable result of our studies is that a large
fraction of the galaxies in our sample do \emph{not} show WR features
and this fraction increases systematically with decreasing
metallicity. We find that 70\% of galaxies at oxygen abundances lower
than 8.2 do not show WR features in their spectra. We discussed a
range of different mechanism responsible for producing
$\heii{\lambda4686}$ line apart from WR stars in these galaxies and
conclude that spatial separation between WR stars and the region
emitting \heii{} emission can be a possible explanation for
non-detection of WR features in these galaxies. Moreover, if the
stellar population models at very low metallicities can have much
higher temperatures than is currently expected in models, as would for
instance be the case if some stars rotate fast enough to evolve
homogeneously, then such models might explain the origin of the
$\heii{\lambda4686}$ line and also the metallicity trend of the
\heii{} sample better. We will explore these possibilities in a future
paper.
 
\section*{Acknowledgements}
First of all we would like to thank the anonymous referee for
insightful comments and valuable suggestions which improved the
paper significantly.

We also would like to thank Marijn Franx, Norbert Langer, S.-C. Yoon
and Brent Groves for useful discussion. Finally, we would like to
express our appreciation to Daniel Schaerer, Johan Eldridge, Carolina
Kehrig and Alireza Rahmati for their kind comments on this paper.

Funding for the Sloan Digital Sky Survey (SDSS) and SDSS-II has been
provided by the Alfred P. Sloan Foundation, the Participating
Institutions, the National Science Foundation, the U.S. Department of
Energy, the National Aeronautics and Space Administration, the
Japanese Monbukagakusho, and the Max Planck Society, and the Higher
Education Funding Council for England. The SDSS Web site is
\texttt{http://www.sdss.org/}.

The SDSS is managed by the Astrophysical Research Consortium (ARC) for
the Participating Institutions. The Participating Institutions are the
American Museum of Natural History, Astrophysical Institute Potsdam,
University of Basel, University of Cambridge, Case Western Reserve
University, The University of Chicago, Drexel University, Fermilab,
the Institute for Advanced Study, the Japan Participation Group, The
Johns Hopkins University, the Joint Institute for Nuclear
Astrophysics, the Kavli Institute for Particle Astrophysics and
Cosmology, the Korean Scientist Group, the Chinese Academy of Sciences
(LAMOST), Los Alamos National Laboratory, the Max-Planck-Institute for
Astronomy (MPIA), the Max-Planck-Institute for Astrophysics (MPA), New
Mexico State University, Ohio State University, University of
Pittsburgh, University of Portsmouth, Princeton University, the United
States Naval Observatory, and the University of Washington.

Many thanks go to Allan Brighton, Thomas Herlin, Miguel Albrecht, 
Daniel Durand and Peter Biereichel, who are responsible for the SkyCat 
developments at ESO, in particular for making their software free for 
general use. GAIA and SkyCat are both based on the scripting language 
Tcl/Tk developed by John Ousterhout and the [incr Tcl] object oriented 
extensions developed by Michael McLennan. They also make use of many 
other extensions and scripts developed by the Tcl community. Thanks 
are also due to the many people who helped test out GAIA and iron out 
minor and major problems (in particular Tim Jenness, Tim Gledhill and 
Nigel Metcalfe) and all the users who have reported bugs and sent support 
since the early releases and continue to do so.
The 3D facilities of GAIA make extensive use of the VTK library. Also a free library.

GAIA was created by the now closed Starlink UK project, funded by the 
Particle Physics and Astronomy Research Council (PPARC) and has been more 
recently supported by the Joint Astronomy Centre Hawaii funded again by 
PPARC and more recently by its successor organisation the Science and 
Technology Facilities Council (STFC). 

This research has made use of the Perl Data Language (PDL,
\texttt{http://pdl.perl.org}) and the Interactive Data Language (IDL).

\setcounter{figure}{17}
\begin{figure*}
  \centerline{\hbox{\includegraphics[trim = 10mm 70mm 10mm 5mm, clip, width=0.75\textwidth]
 {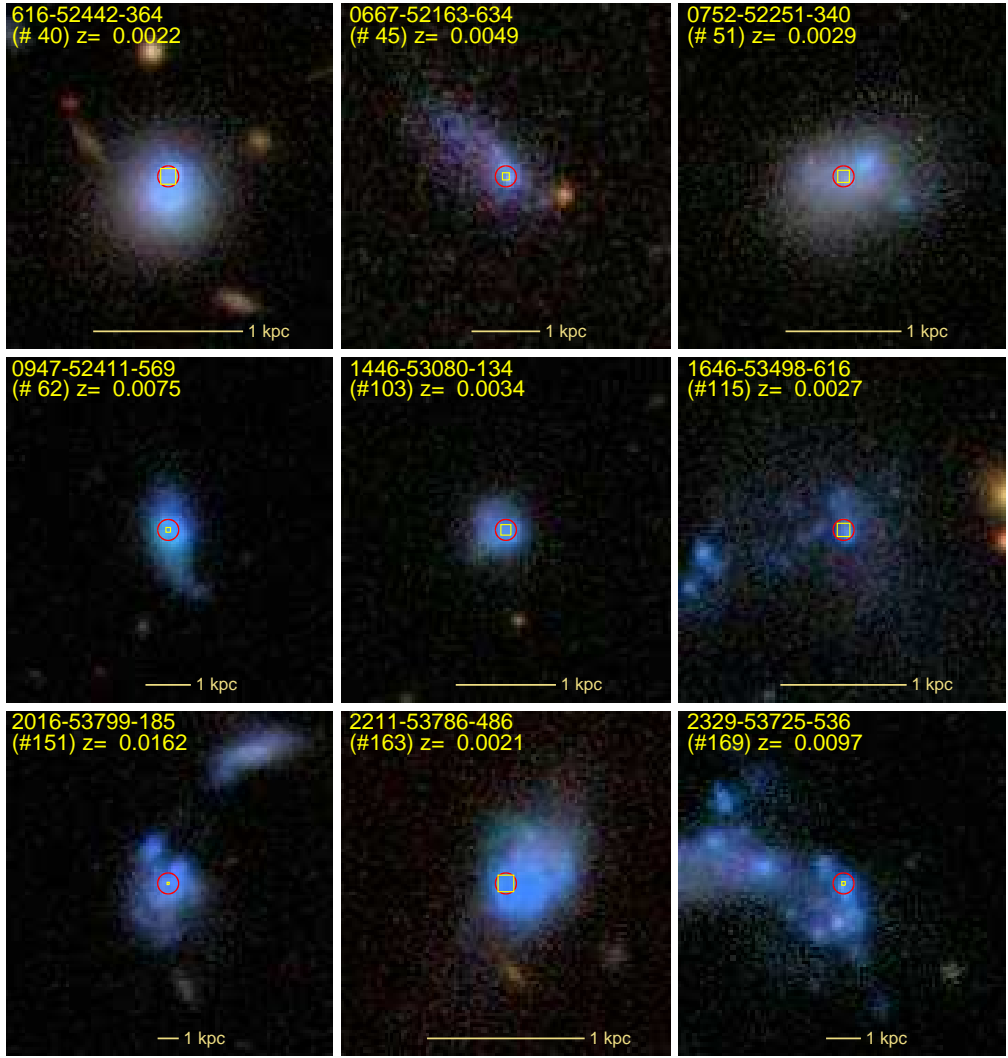}}} 
\caption{Images of nine galaxies which do not show WR features in
  their spectra while have strong $\heii{\lambda4686}$ emission.  The
  sizes of boxes are 50 '' $\times$ 50 ''. In these images north is
  up, east to the left. Red circles show the size of SDSS fibre
  aperture (3'') and yellow boxes show 100 pc around the center of the
  SDSS fibre.}
\label{fig18}
\end{figure*}


\begin{appendix}

\section{Fitting models to the emission lines}
The CL01 model grid is calculated by varying the model parameters,
$U$, ionization parameter at the edge of the Str{\"o}mgren sphere,
$\tau_V$, total $V$--band optical depth, $\xi$, the dust-to-metal
ratio of ionized gas, $\mu$, the fraction of the total optical depth
in the neutral ISM contributed by the ambient ISM, and $Z$, the
metallicity, over a certain range for 221 unequally spaced time
steps from $t=0$ to $t=20 Gyr$ (see Table~\ref{table2}).\\
In this paper we use the interpolated model grid of various
luminosities for 50 time-steps from B04.  This includes a total of
$2\times10^5$ different models. To fit to the data we adopt the
Bayesian methodology described by Ka03.  We obtain the PDF of every
parameter of interest by marginalisation over all other
parameters. The resulting PDF is used to estimate confidence intervals
for each estimated physical parameter.  We need to fit to at least
five strong emission lines, $\oii{\lambda3727,3729}$, $\hb$,
$\oiii{\lambda5007}$, $\ha$, $\nii{\lambda6584}$ to get a good
constraint on the parameters. We take the median value of each
parameter to be the best estimate of a given parameter. 

\begin{figure*}
\centerline{{\includegraphics[width=0.5\textwidth, angle=90]
           {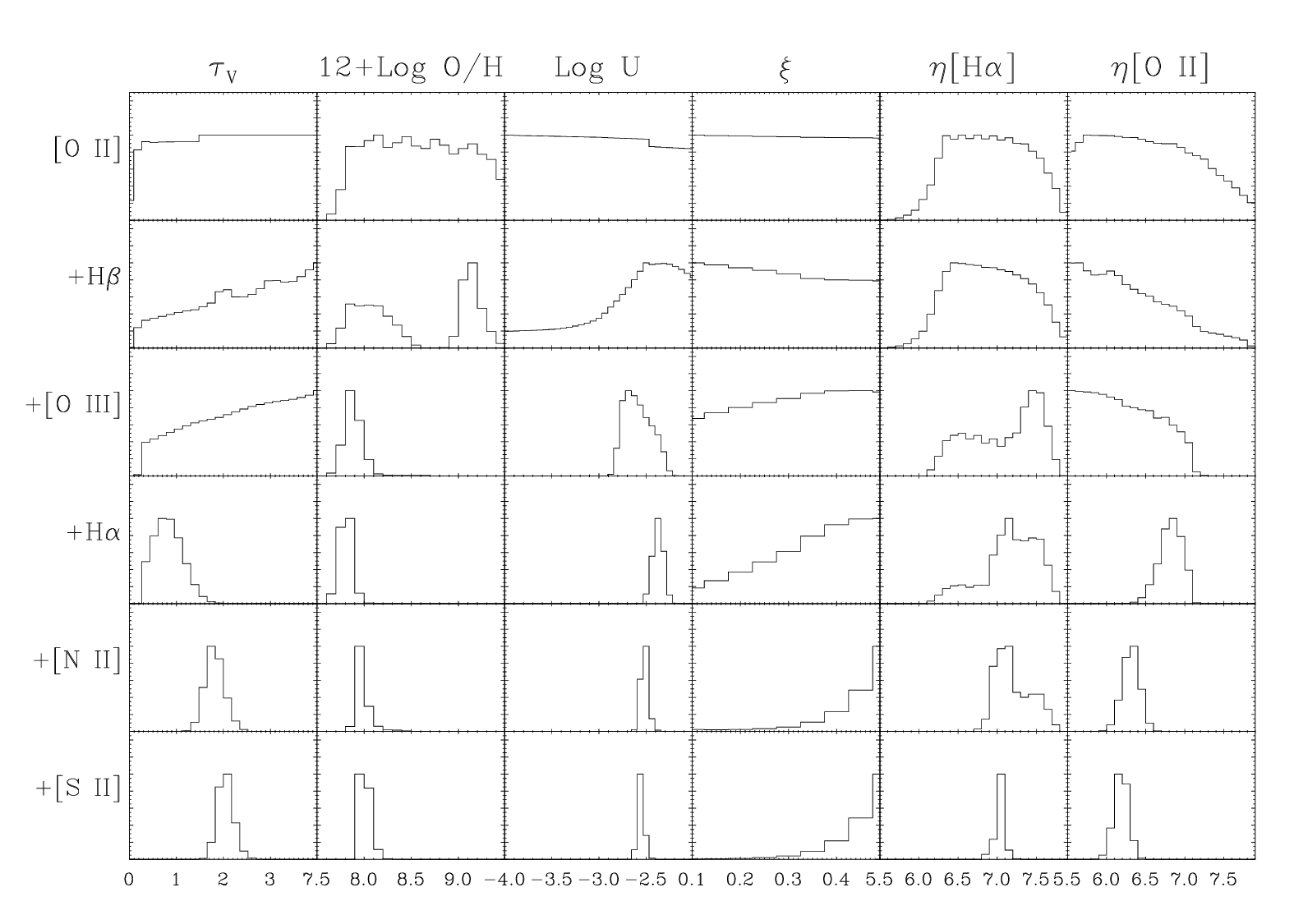}}}
       \caption{ We show how the PDFs for dust attenuation parameter
         in $V$-band, gas phase oxygen abundance, ionization parameter
         of ionized gas, dust-to-metal ratio and \ha\ and \oii\
         efficiency factors change when we add more emission lines to
         the fit. $\xi$ is hard to constrain, except at high
         metallicity.}
\label{fig 26}
\end{figure*}

In Figure~\ref{fig 26} we illustrate our technique by showing the
effect of adding lines on the PDFs of parameters when we fit a model
to the data.  We start with $\oii{\lambda3727,3729}$ and show how we
get more well defined PDFs for the indicated parameters as we add the
emission lines indicated on the left.  We show the PDFs
for dust attenuation parameter in $V$ -band, gas phase oxygen
abundance, ionization parameter, dust-to-metal ratio of ionized gas
and the conversion factor from \ha\ and \oii\ luminosity to star
formation rate (see CL01 for further details), for one object in our
sample. The dust-to-metal ratio, $\xi$, is hard to constrain, except
at high metallicity.
\end{appendix}

\clearpage

\setcounter{table}{1}
\onecolumn
\begin{landscape}
\begin{longtable}{cccccllcll}
\caption{\label{table5} The positions and the identifications of galaxies with nebular $\heii{\lambda4686}$ emission. 
OKSN is for galaxies having $SN >3$ in their $\oiii{\lambda5007}$, \hb, $\nii{\lambda6584}$ and \ha\ lines. 
See B04 for the BPT classification. The full table is available in electronic form in http://www.strw.leidenuniv.nl/$\sim$shirazi/SB011/.}\\
 \hline
\multicolumn{1}{c}{$\alpha(J2000)$} & \multicolumn{1}{c}{$\delta$} &
\multicolumn{1}{c}{Plate--MJD--FiberID} &
\multicolumn{1}{l}{$\log \heii/\hb$} & \multicolumn{1}{l}{$\log \nii/\ha$} & 
\multicolumn{1}{l}{Class} & \multicolumn{1}{l}{Class(BPT)} & 
\multicolumn{1}{c}{SN} & WR Features & Class(WR)\\
\hline
+11 59 09.71 & +00 00 06.78 & 0285-51930-485 & -0.93 & -0.14 & AGN & AGN & OKSN & Non-WR & --- \\
+15 30 26.32 & +00 00 10.81 & 0363-51989-400 & -0.79 & -0.31 & AGN & AGN & OKSN & Non-WR & --- \\
+08 58 28.60 & +00 01 24.49 & 0470-51929-309 & -0.84 & -0.01 & AGN & AGN & OKSN & WR &  1 \\
+13 06 00.68 & +00 01 25.06 & 0294-51986-438 & -1.12 & -0.07 & AGN & AGN & OKSN & Non-WR &  0 \\
+01 10 06.09 & +00 01 33.00 & 0694-52209-113 & -0.88 & -0.36 & AGN & AGN & OKSN & Non-WR & --- \\
\hline 
\end{longtable}

\setcounter{table}{4}
\begin{longtable}{lcccccccclll}
\caption{\label{table4} The positions and identifications of the sample of star-forming 
galaxies with nebular $\heii{\lambda4686}$ emission.}\\
 \hline
\multicolumn{1}{l}{Plate--MJD--FiberID} & \multicolumn{1}{c}{Z} &
\multicolumn{1}{c}{$\alpha(J2000)$} & \multicolumn{1}{c}{$\delta$} &
\multicolumn{1}{c}{ID} &
\multicolumn{1}{l}{$\log \heii/\hb$} & \multicolumn{1}{l}{$\log \nii/\ha$} &
\multicolumn{1}{l}{$12+\log \rm{O/H}$} & Features & Class & Other Names\\
\hline
\endfirsthead
\caption[]{continued.}\\
\hline
\multicolumn{1}{l}{Plate--MJD--FiberID} & \multicolumn{1}{c}{Z} &
\multicolumn{1}{c}{$\alpha(J2000)$} & \multicolumn{1}{c}{$\delta$} &
\multicolumn{1}{c}{ID} & 
\multicolumn{1}{l}{$\log \heii/\hb$}  & \multicolumn{1}{l}{$\log \nii/\ha$} &
\multicolumn{1}{l}{$12+\log \rm{O/H}$} & Features & Class & Other Names\\
\hline
\endhead
\hline
0752-52251-340 & 0.00291 & +00 09 53.09 & +15 44 04.80 &  51 & -1.35 & -1.67 & 7.92$\pm$0.09 & Non-WR & 0 & N/A  \\
0390-51900-291 & 0.09437 & +00 17 28.29 & -00 56 24.98 &  13 & -1.57 & -0.56 & 9.27$\pm$0.05 & WR & 2 & N/A  \\
0390-51900-445 & 0.09840 & +00 21 01.03 & +00 52 48.08 &  14 & -1.96 & -1.10 & 8.83$\pm$0.04 & WR & 2 & UM 228 \\
0753-52233-094 & 0.01424 & +00 24 25.95 & +14 04 10.65 &  52 & -1.98 & -1.36 & 8.10$\pm$0.12 & WR & 3 & N/A  \\
0418-51884-319 & 0.01791 & +00 32 18.59 & +15 00 14.17 &  19 & -1.97 & -1.52 & 8.01$\pm$0.07 & WR & 2 & SHOC 022 \\
0691-52199-389 & 0.19554 & +00 42 52.31 & +00 27 30.09 &  48 & -0.75 & -0.82 & 8.88$\pm$0.05 & WR & 2 & N/A  \\
1905-53706-628 & 0.15928 & +00 49 13.88 & +00 24 01.99 & 138 & -1.64 & -2.05 & 7.85$\pm$0.06 & Non-WR & 0 & N/A  \\
0394-51913-075 & 0.16742 & +00 55 27.46 & -00 21 48.77 &  15 & -1.76 & -0.89 & 8.90$\pm$0.04 & WR & 2 & N/A  \\
0695-52202-137 & 0.00556 & +01 15 33.82 & -00 51 31.17 &  49 & -2.32 & -1.27 & 8.31$\pm$0.35 & WR & 3 & NGC 0450 \\
2329-53725-536 & 0.00975 & +01 25 34.19 & +07 59 24.40 & 169 & -2.02 & -2.22 & 7.80$\pm$0.04 & Non-WR & 0 & N/A  \\
0429-51820-495 & 0.05662 & +01 47 07.04 & +13 56 29.29 &  20 & -1.95 & -1.52 & 8.02$\pm$0.05 & Non-WR & 0 & N/A  \\
0666-52149-331 & 0.15155 & +01 59 53.07 & -08 13 48.99 &  44 & -1.73 & -1.43 & 8.04$\pm$0.06 & Non-WR & 0 & SHOC 099 \\
1073-52649-409 & 0.03994 & +02 13 06.62 & +00 56 12.44 &  73 & -1.97 & -1.58 & 8.09$\pm$0.05 & Non-WR & 0 & UM 411 \\
0667-52163-634 & 0.00492 & +02 15 13.98 & -08 46 24.39 &  45 & -1.12 & -1.81 & 7.74$\pm$0.09 & Non-WR & 0 & SHOC 111 \\
0456-51910-306 & 0.08222 & +02 40 52.20 & -08 28 27.43 &  24 & -1.60 & -1.63 & 8.02$\pm$0.04 & Non-WR & 0 & SHOC 133 \\
1070-52591-072 & 0.00421 & +02 42 39.86 & -00 00 58.64 &  72 & -1.41 & -0.33 & 8.93$\pm$0.25 & Non-WR & 0 & M077  \\
0456-51910-076 & 0.00456 & +02 48 15.94 & -08 17 16.51 &  23 & -2.11 & -1.90 & 7.87$\pm$0.08 & Non-WR & 0 & N/A  \\
1666-52991-310 & 0.03992 & +03 14 31.71 & +41 05 25.95 & 117 & -1.95 & -1.40 & 8.08$\pm$0.19 & WR & 1 & N/A  \\
0413-51821-480 & 0.06868 & +03 17 43.12 & +00 19 36.84 &  16 & -0.77 & -0.54 & 9.07$\pm$0.11 & WR & 2 & N/A  \\
1733-53047-528 & 0.07881 & +07 29 30.29 & +39 49 41.62 & 121 & -1.97 & -0.93 & 8.90$\pm$0.04 & Non-WR & 0 & N/A  \\
1922-53315-588 & 0.06978 & +08 06 19.50 & +19 49 27.31 & 139 & -2.05 & -1.19 & 8.83$\pm$0.04 & WR & 2 & N/A  \\
0761-52266-361 & 0.04592 & +08 22 27.44 & +42 23 31.14 &  53 & -1.09 & -0.37 & 9.25$\pm$0.07 & WR & 1 & N/A  \\
1267-52932-384 & 0.04722 & +08 23 54.97 & +28 06 21.75 &  78 & -2.16 & -0.95 & 8.88$\pm$0.04 & WR & 1 & N/A  \\
0548-51986-503 & 0.00706 & +08 26 04.80 & +45 58 07.36 &  30 & -2.07 & -0.76 & 8.64$\pm$0.26 & WR & 1 & UGC 04393 \\
2425-54139-372 & 0.01974 & +08 29 32.66 & +14 27 06.92 & 173 & -2.00 & -1.35 & 8.13$\pm$0.20 & Non-WR & 0 & N/A  \\
0445-51873-404 & 0.00245 & +08 37 43.48 & +51 38 30.26 &  21 & -2.38 & -1.71 & 7.95$\pm$0.08 & Non-WR & 0 & MRK 0094 \\
0828-52317-148 & 0.14746 & +08 38 43.64 & +38 53 50.50 &  56 & -1.88 & -1.32 & 8.69$\pm$0.05 & WR & 1 & N/A  \\
2278-53711-411 & 0.07219 & +08 40 00.37 & +18 05 31.01 & 168 & -2.02 & -1.53 & 7.98$\pm$0.07 & Non-WR & 0 & N/A  \\
0564-52224-216 & 0.09109 & +08 44 14.24 & +02 26 21.10 &  34 & -2.13 & -1.20 & 8.75$\pm$0.13 & WR & 1 & N/A  \\
1875-54453-549 & 0.07498 & +08 51 03.67 & +62 13 26.93 & 136 & -2.06 & -1.39 & 8.59$\pm$0.10 & Non-WR & 0 & N/A  \\
1785-54439-201 & 0.09186 & +08 51 15.65 & +58 40 55.02 & 130 & -2.01 & -1.83 & 7.84$\pm$0.08 & Non-WR & 0 & N/A  \\
2430-53815-117 & 0.07592 & +08 52 21.72 & +12 16 51.76 & 174 & -2.06 & -1.45 & 8.23$\pm$0.08 & Non-WR & 0 & N/A  \\
0551-51993-279 & 0.00961 & +08 52 58.21 & +49 27 33.91 &  31 & -1.83 & -0.62 & 8.80$\pm$0.24 & WR & 3 & SBS 0849+496 \\
2086-53401-458 & 0.00981 & +09 05 26.34 & +25 33 02.57 & 154 & -1.93 & -1.12 & 8.40$\pm$0.45 & WR & 3 & UGC 04764 \\
0566-52238-497 & 0.03909 & +09 05 31.08 & +03 35 30.38 &  35 & -1.93 & -1.78 & 7.79$\pm$0.08 & Non-WR & 0 & N/A  \\
1194-52703-397 & 0.00513 & +09 10 28.78 & +07 11 17.97 &  76 & -2.02 & -1.77 & 7.91$\pm$0.09 & WR & 1 & N/A  \\
0899-52620-594 & 0.02727 & +09 14 34.95 & +47 02 07.24 &  60 & -2.13 & -1.55 & 7.97$\pm$0.07 & Non-WR & 0 & SBS 0911+472 \\
0554-52000-190 & 0.00767 & +09 20 55.92 & +52 34 07.34 &  33 & -2.10 & -1.83 & 7.85$\pm$0.08 & Non-WR & 0 & N/A  \\
0553-51999-602 & 0.00772 & +09 20 56.07 & +52 34 04.32 &  32 & -1.71 & -1.63 & 7.88$\pm$0.07 & WR & 1 & N/A  \\
0485-51909-550 & 0.01366 & +09 30 06.43 & +60 26 53.40 &  25 & -2.17 & -1.59 & 7.98$\pm$0.06 & Non-WR & 0 & SBS 0926+606A \\
2580-54092-470 & 0.10060 & +09 38 01.64 & +13 53 17.07 & 186 & -1.18 & -0.43 & 9.28$\pm$0.06 & WR & 2 & N/A  \\
1594-52992-563 & 0.01486 & +09 42 52.78 & +35 47 25.98 & 110 & -1.92 & -1.44 & 8.19$\pm$0.18 & Non-WR & 0 & N/A  \\
1305-52757-269 & 0.01085 & +09 42 56.74 & +09 28 16.26 &  80 & -2.38 & -1.51 & 8.14$\pm$0.16 & WR & 3 & UGC 05189 \\
0266-51630-100 & 0.00478 & +09 44 01.87 & -00 38 32.18 &   2 & -1.88 & -1.98 & 7.77$\pm$0.06 & Non-WR & 0 & CGCG 007-025 \\
1947-53431-448 & 0.01730 & +09 50 00.77 & +30 03 41.04 & 140 & -1.97 & -1.51 & 8.04$\pm$0.10 & Non-WR & 0 & N/A  \\
0769-54530-086 & 0.04626 & +09 51 31.77 & +52 59 36.05 &  54 & -2.05 & -1.64 & 7.93$\pm$0.06 & WR & 2 & SBS 0948+532 \\
1306-52996-005 & 0.00486 & +09 54 49.56 & +09 16 15.94 &  81 & -1.90 & -0.42 & 9.03$\pm$0.21 & WR & 3 & NGC 3049 \\
2583-54095-062 & 0.01522 & +09 56 42.49 & +15 38 11.34 & 187 & -2.02 & -0.91 & 8.56$\pm$0.28 & WR & 3 & UGC 05342 \\
2364-53737-618 & 0.00423 & +10 10 32.81 & +22 00 39.63 & 171 & -2.12 & -1.91 & 7.82$\pm$0.09 & WR & 1 & WAS 05 \\
1745-53061-475 & 0.06131 & +10 10 42.54 & +12 55 16.81 & 123 & -1.92 & -1.40 & 8.40$\pm$0.04 & WR & 3 & N/A  \\
2588-54174-369 & 0.05563 & +10 10 59.30 & +15 42 23.53 & 188 & -1.92 & -1.70 & 7.91$\pm$0.04 & Non-WR & 0 & N/A  \\
1745-53061-196 & 0.00957 & +10 12 27.02 & +12 20 37.50 & 122 & -2.24 & -1.77 & 8.04$\pm$0.10 & WR & 2 & N/A  \\
1955-53442-354 & 0.03762 & +10 14 10.58 & +34 20 34.79 & 141 & -1.66 & -0.54 & 9.13$\pm$0.05 & WR & 1 & KUG 1011+345 \\
1427-52996-221 & 0.00388 & +10 16 24.52 & +37 54 45.97 & 101 & -1.70 & -2.21 & 7.80$\pm$0.05 & Non-WR & 0 & N/A  \\
2366-53741-124 & 0.01897 & +10 24 02.75 & +21 04 50.04 & 172 & -1.88 & -1.56 & 8.08$\pm$0.09 & WR & 2 & LSBC D568-03 \\
0575-52319-521 & 0.03319 & +10 24 29.25 & +05 24 51.02 &  36 & -2.07 & -1.72 & 7.84$\pm$0.07 & Non-WR & 0 & N/A  \\
2591-54140-222 & 0.04432 & +10 26 23.65 & +17 10 14.36 & 189 & -1.12 & -0.64 & 9.12$\pm$0.05 & Non-WR & 0 & N/A  \\
0999-52636-517 & 0.04453 & +10 33 28.53 & +07 08 01.76 &  69 & -1.99 & -0.77 & 8.91$\pm$0.04 & WR & 2 & CGCG 037-076 \\
0947-52411-569 & 0.00747 & +10 34 10.15 & +58 03 49.06 &  62 & -1.60 & -1.97 & 7.78$\pm$0.06 & Non-WR & 0 & MRK 1434 \\
0875-52354-226 & 0.02850 & +10 35 08.88 & +49 21 42.47 &  58 & -1.79 & -0.76 & 8.92$\pm$0.04 & WR & 2 & SBS 1032+496 \\
1998-53433-304 & 0.00489 & +10 36 13.22 & +37 19 27.57 & 146 & -1.93 & -0.53 & 9.09$\pm$0.26 & WR & 3 & NGC 3294 \\
0274-51913-187 & 0.01856 & +10 39 24.38 & -00 23 21.44 &   3 & -2.16 & -0.85 & 8.67$\pm$0.26 & WR & 1 & IC 0633 \\
2478-54097-370 & 0.00398 & +10 41 09.60 & +21 21 42.80 & 176 & -2.13 & -1.81 & 7.83$\pm$0.09 & WR & 3 & MRK 0724 \\
0578-52339-060 & 0.01287 & +10 44 57.79 & +03 53 13.15 &  37 & -1.80 & -2.55 & 7.80$\pm$0.00 & Non-WR & 0 & N/A  \\
2147-53491-514 & 0.05487 & +10 45 20.42 & +09 23 49.10 & 160 & -2.18 & -1.30 & 8.70$\pm$0.06 & WR & 2 & SCHG 1042+097 \\
0275-51910-445 & 0.02620 & +10 45 54.78 & +01 04 05.84 &   4 & -2.28 & -1.34 & 8.74$\pm$0.04 & Non-WR & 0 & SHOC 308 \\
1749-53357-499 & 0.01061 & +10 46 53.99 & +13 46 45.77 & 124 & -2.23 & -1.80 & 7.92$\pm$0.09 & Non-WR & 0 & N/A  \\
1981-53463-438 & 0.02945 & +10 47 23.61 & +30 21 44.29 & 143 & -2.18 & -1.15 & 8.85$\pm$0.04 & WR & 3 & TON 0542 \\
2483-53852-254 & 0.08443 & +10 50 32.51 & +15 38 06.31 & 177 & -1.79 & -1.76 & 7.92$\pm$0.04 & Non-WR & 0 & N/A  \\
0275-51910-622 & 0.05061 & +10 50 46.59 & +00 36 40.11 &   5 & -1.44 & -0.46 & 9.16$\pm$0.26 & WR & 2 & N/A  \\
0876-52669-175 & 0.00435 & +10 53 10.82 & +50 16 53.21 &  59 & -1.70 & -1.75 & 7.94$\pm$0.08 & Non-WR & 0 & N/A  \\
2359-53826-205 & 0.00452 & +10 54 21.87 & +27 14 22.16 & 170 & -1.93 & -0.74 & 8.67$\pm$0.41 & WR & 3 & NGC 3451 \\
1362-53050-617 & 0.03745 & +11 00 24.90 & +43 01 11.93 &  91 & -2.11 & -1.29 & 8.77$\pm$0.08 & Non-WR & 0 & N/A  \\
2213-53792-359 & 0.00214 & +11 04 58.30 & +29 08 16.55 & 164 & -1.73 & -1.84 & 7.79$\pm$0.05 & Non-WR & 0 & N/A  \\
2211-53786-486 & 0.00214 & +11 04 58.54 & +29 08 15.72 & 163 & -2.18 & -1.83 & 7.81$\pm$0.06 & Non-WR & 0 & N/A  \\
1363-53053-510 & 0.02154 & +11 05 08.12 & +44 44 47.24 &  92 & -2.05 & -1.28 & 8.78$\pm$0.04 & WR & 2 & MRK 0162 \\
2494-54174-361 & 0.00492 & +11 17 46.30 & +17 44 24.69 & 178 & -2.04 & -1.69 & 7.90$\pm$0.09 & Non-WR & 0 & N/A  \\
1223-52781-128 & 0.00345 & +11 27 10.93 & +08 43 51.70 &  77 & -1.71 & -1.27 & 8.09$\pm$0.11 & WR & 3 & IC 2828 \\
1014-52707-254 & 0.00988 & +11 27 32.67 & +53 54 54.47 &  70 & -1.98 & -1.62 & 8.04$\pm$0.11 & Non-WR & 0 & MRK 1446 \\
2500-54178-084 & 0.00470 & +11 29 14.15 & +20 34 52.01 & 179 & -2.32 & -1.44 & 8.03$\pm$0.10 & WR & 2 & IC 0700 \\
1754-53385-151 & 0.01764 & +11 32 35.35 & +14 11 29.83 & 125 & -2.19 & -1.48 & 8.20$\pm$0.09 & WR & 2 & N/A  \\
0967-52636-302 & 0.00084 & +11 33 28.95 & +49 14 13.01 &  64 & -1.22 & -1.76 & 7.89$\pm$0.08 & WR & 3 & Mrk 0178 \\
0967-52636-339 & 0.02601 & +11 34 45.72 & +50 06 03.33 &  65 & -2.25 & -1.41 & 8.21$\pm$0.54 & WR & 2 & MRK 1448 \\
1442-53050-599 & 0.01018 & +11 36 23.82 & +47 09 29.08 & 102 & -2.07 & -1.92 & 7.83$\pm$0.09 & Non-WR & 0 & N/A  \\
2012-53493-407 & 0.00483 & +11 36 39.57 & +36 23 42.89 & 150 & -1.82 & -1.52 & 8.07$\pm$0.10 & WR & 1 & NGC 3755 \\
2506-54179-357 & 0.02082 & +11 36 54.01 & +19 55 34.80 & 180 & -1.77 & -0.77 & 8.61$\pm$0.40 & WR & 3 & MRK 0182 \\
2008-53473-467 & 0.00601 & +11 41 07.49 & +32 25 37.22 & 149 & -1.82 & -1.82 & 7.79$\pm$0.05 & Non-WR & 0 & KUG 1138+327 \\
0967-52636-540 & 0.00558 & +11 45 06.26 & +50 18 02.44 &  66 & -1.55 & -2.04 & 7.77$\pm$0.09 & Non-WR & 0 & N/A  \\
2513-54141-309 & 0.02676 & +11 48 05.45 & +21 49 45.35 & 183 & -1.95 & -1.27 & 8.35$\pm$0.11 & WR & 2 & MRK 1459 \\
2510-53877-560 & 0.04512 & +11 48 27.34 & +25 46 11.77 & 182 & -2.10 & -1.70 & 7.95$\pm$0.07 & Non-WR & 0 & N/A  \\
2508-53875-615 & 0.07911 & +11 48 40.87 & +17 56 33.02 & 181 & -2.00 & -1.78 & 7.91$\pm$0.04 & Non-WR & 0 & N/A  \\
1761-53376-636 & 0.00245 & +11 50 02.73 & +15 01 23.48 & 126 & -2.23 & -1.59 & 7.99$\pm$0.05 & WR & 3 & MRK 0750 \\
0330-52370-471 & 0.00353 & +11 52 37.68 & -02 28 06.39 &   8 & -2.13 & -1.75 & 7.84$\pm$0.09 & Non-WR & 0 & N/A  \\
1446-53080-134 & 0.00337 & +11 54 41.22 & +46 36 36.35 & 103 & -1.81 & -1.68 & 7.80$\pm$0.05 & Non-WR & 0 & N/A  \\
1313-52790-423 & 0.01726 & +11 55 28.34 & +57 39 51.97 &  82 & -2.20 & -1.77 & 8.03$\pm$0.10 & Non-WR & 0 & MRK 0193 \\
0516-52017-315 & 0.05811 & +11 57 12.45 & +02 28 27.88 &  27 & -1.99 & -1.15 & 8.85$\pm$0.06 & WR & 1 & UM 469 \\
1991-53446-584 & 0.01097 & +11 57 31.73 & +32 20 30.17 & 145 & -2.14 & -1.13 & 8.36$\pm$0.49 & WR & 3 & NGC 3991N \\
2226-53819-157 & 0.08188 & +12 00 16.49 & +27 19 59.01 & 165 & -1.93 & -1.86 & 7.94$\pm$0.07 & Non-WR & 0 & N/A  \\
1763-53463-094 & 0.06675 & +12 00 33.42 & +13 43 07.99 & 127 & -2.05 & -1.08 & 8.72$\pm$0.04 & WR & 2 & N/A  \\
2227-53820-389 & 0.05588 & +12 01 49.90 & +28 06 10.67 & 166 & -1.92 & -1.71 & 7.92$\pm$0.05 & Non-WR & 0 & N/A  \\
0517-52024-504 & 0.00429 & +12 08 11.11 & +02 52 41.82 &  28 & -1.84 & -0.27 & 9.04$\pm$0.17 & WR & 2 & NGC 4123 \\
2004-53737-439 & 0.04894 & +12 09 24.64 & +32 44 02.05 & 148 & -1.44 & -0.43 & 9.27$\pm$0.04 & WR & 1 & N/A  \\
2644-54210-188 & 0.02435 & +12 09 27.95 & +22 06 16.69 & 193 & -1.91 & -0.64 & 9.13$\pm$0.08 & WR & 2 & UGC 07137 \\
2610-54476-421 & 0.00277 & +12 15 18.60 & +20 38 26.72 & 191 & -2.59 & -1.80 & 7.88$\pm$0.08 & WR & 2 & MRK 1315 \\
1625-53140-386 & 0.00864 & +12 16 47.89 & +08 02 56.28 & 112 & -1.81 & -1.44 & 7.99$\pm$0.05 & Non-WR & 0 & VCC 0207 \\
2001-53493-146 & 0.00061 & +12 17 49.31 & +37 51 55.50 & 147 & -2.28 & -1.72 & 7.92$\pm$0.09 & Non-WR & 0 & N/A  \\
2880-54509-277 & 0.00428 & +12 22 25.79 & +04 34 04.77 & 198 & -2.28 & -1.33 & 8.10$\pm$0.11 & Non-WR & 0 & N/A  \\
0955-52409-608 & 0.00234 & +12 25 05.41 & +61 09 11.29 &  63 & -1.97 & -1.96 & 7.96$\pm$0.07 & WR & 3 & SBS 1222+614 \\
2880-54509-095 & 0.09423 & +12 26 11.90 & +04 15 36.06 & 197 & -2.00 & -1.66 & 7.96$\pm$0.06 & Non-WR & 0 & N/A  \\
1453-53084-322 & 0.00100 & +12 26 15.69 & +48 29 38.43 & 106 & -2.27 & -2.11 & 7.80$\pm$0.04 & Non-WR & 0 & UGCA 281 \\
1371-52821-053 & 0.00071 & +12 28 09.26 & +44 05 08.02 &  93 & -1.67 & -1.09 & 8.22$\pm$0.75 & WR & 1 & NGC 4449 \\
1371-52821-059 & 0.00072 & +12 28 13.86 & +44 07 10.43 &  94 & -2.50 & -1.95 & 7.63$\pm$0.06 & WR & 3 & NGC 4449 \\
1452-53112-011 & 0.00157 & +12 30 28.33 & +41 41 22.07 & 104 & -2.33 & -1.23 & 8.29$\pm$0.43 & WR & 3 & NGC 4485 \\
1452-53112-016 & 0.00192 & +12 30 38.45 & +41 39 11.31 & 105 & -2.14 & -1.31 & 8.07$\pm$0.08 & WR & 3 & NGC 4490 \\
1615-53166-120 & 0.00418 & +12 30 48.60 & +12 02 42.82 & 111 & -1.44 & -2.04 & 7.73$\pm$0.07 & WR & 1 & N/A  \\
1768-53442-476 & 0.00697 & +12 31 54.67 & +15 07 36.48 & 128 & -1.82 & -0.68 & 9.18$\pm$0.10 & WR & 3 & IC 0797 \\
2613-54481-507 & 0.04855 & +12 38 29.93 & +19 59 21.36 & 192 & -1.90 & -0.99 & 8.93$\pm$0.04 & WR & 2 & N/A  \\
0494-51915-007 & 0.08778 & +12 40 49.89 & +66 24 20.17 &  26 & -1.88 & -1.49 & 7.91$\pm$0.07 & Non-WR & 0 & SHOC 379 \\
1372-53062-072 & 0.04190 & +12 41 34.25 & +44 26 39.24 &  95 & -0.19 & -0.93 & 8.80$\pm$0.08 & Non-WR & 0 & N/A  \\
1975-53734-498 & 0.00198 & +12 43 56.70 & +32 10 14.67 & 142 & -1.77 & -1.88 & 7.93$\pm$0.08 & WR & 3 & NGC 4656 \\
2236-53729-038 & 0.00350 & +12 45 16.87 & +27 07 30.78 & 167 & -1.83 & -1.15 & 8.15$\pm$0.19 & WR & 3 & NGC 4670 \\
1455-53089-556 & 0.02377 & +12 48 46.36 & +47 42 53.45 & 107 & -2.14 & -1.48 & 8.13$\pm$0.06 & WR & 2 & N/A  \\
1989-53772-089 & 0.02782 & +12 53 06.56 & +36 49 11.41 & 144 & -1.85 & -0.52 & 9.27$\pm$0.07 & WR & 3 & NGC 4774 \\
2461-54570-089 & 0.00797 & +12 54 23.74 & +58 53 40.67 & 175 & -2.03 & -1.15 & 8.23$\pm$0.62 & WR & 3 & N/A  \\
0602-52072-369 & 0.02766 & +13 02 49.20 & +65 34 49.27 &  39 & -1.92 & -1.57 & 8.04$\pm$0.06 & WR & 1 & N/A  \\
2018-53800-096 & 0.03595 & +13 03 54.44 & +37 14 01.88 & 152 & -2.11 & -1.39 & 8.54$\pm$0.11 & Non-WR & 0 & N/A  \\
0339-51692-083 & 0.00450 & +13 04 32.27 & -03 33 22.12 &   9 & -2.26 & -1.36 & 8.10$\pm$0.09 & WR & 3 & UGCA 322 \\
0782-52320-022 & 0.11183 & +13 04 45.63 & +62 24 20.88 &  55 & -1.84 & -0.95 & 8.94$\pm$0.04 & WR & 2 & N/A  \\
2016-53799-185 & 0.01620 & +13 06 24.19 & +35 13 43.04 & 151 & -2.08 & -1.71 & 7.87$\pm$0.08 & Non-WR & 0 & N/A  \\
0602-52072-019 & 0.03678 & +13 14 26.56 & +63 33 11.37 &  38 & -1.62 & -1.15 & 8.87$\pm$0.08 & WR & 3 & N/A  \\
2023-53851-263 & 0.00288 & +13 14 47.37 & +34 52 59.81 & 153 & -2.37 & -1.45 & 8.22$\pm$0.20 & WR & 3 & UGC 08323 \\
0526-52312-097 & 0.18400 & +13 22 11.96 & +01 30 34.39 &  29 & -2.04 & -1.00 & 8.89$\pm$0.05 & WR & 2 & F13196+0146  \\
0341-51690-606 & 0.02246 & +13 23 47.46 & -01 32 51.95 &  10 & -1.88 & -2.48 & 7.80$\pm$0.03 & Non-WR & 0 & UM 570 \\
1282-52759-057 & 0.01637 & +13 25 19.89 & +48 02 26.15 &  79 & -1.93 & -1.64 & 7.94$\pm$0.08 & Non-WR & 0 & SBS 1323+483 \\
2112-53534-557 & 0.01462 & +13 25 49.42 & +33 03 54.38 & 157 & -2.05 & -1.57 & 8.15$\pm$0.16 & WR & 2 & WAS 69 \\
1376-53089-637 & 0.02797 & +13 28 44.05 & +43 55 50.51 &  96 & -2.16 & -1.28 & 8.84$\pm$0.06 & WR & 2 & MRK 0259 \\
2606-54154-474 & 0.09425 & +13 29 16.56 & +17 00 21.00 & 190 & -1.92 & -1.13 & 8.80$\pm$0.09 & WR & 3 & N/A  \\
2110-53467-499 & 0.01613 & +13 30 17.38 & +31 19 58.02 & 156 & -2.14 & -1.41 & 8.11$\pm$0.14 & Non-WR & 0 & UGC 08496 \\
1464-53091-370 & 0.01170 & +13 31 26.91 & +41 51 48.29 & 108 & -2.13 & -2.12 & 7.79$\pm$0.04 & Non-WR & 0 & N/A  \\
1801-54156-583 & 0.14730 & +13 39 24.24 & +07 39 27.61 & 131 & -1.80 & -0.82 & 8.91$\pm$0.04 & WR & 1 & N/A  \\
2094-53851-487 & 0.00320 & +13 41 56.48 & +30 31 09.62 & 155 & -2.19 & -1.73 & 7.98$\pm$0.07 & WR & 2 & Mrk 0067c3 \\
1043-52465-308 & 0.00594 & +13 42 51.85 & +52 42 30.57 &  71 & -1.79 & -1.53 & 7.99$\pm$0.06 & WR & 1 & MRK 1480 \\
1803-54152-448 & 0.05450 & +13 44 24.06 & +07 45 00.08 & 132 & -1.78 & -1.00 & 8.93$\pm$0.04 & WR & 2 & N/A  \\
0854-52373-514 & 0.03041 & +13 45 31.50 & +04 42 32.71 &  57 & -1.97 & -1.55 & 7.92$\pm$0.04 & WR & 1 & TOLOLO 1343+049 \\
1776-53858-632 & 0.02156 & +13 46 49.45 & +14 24 01.68 & 129 & -1.75 & -0.42 & 8.92$\pm$0.32 & WR & 2 & MRK 0796 \\
1158-52668-062 & 0.03383 & +13 59 50.92 & +57 26 22.98 &  75 & -1.89 & -1.42 & 8.03$\pm$0.08 & Non-WR & 0 & MRK 1486 \\
2770-54510-583 & 0.02775 & +14 01 07.12 & +21 14 34.60 & 195 & -2.14 & -1.20 & 8.86$\pm$0.04 & WR & 1 & UGC 08929 \\
1324-53088-271 & 0.00058 & +14 02 28.23 & +54 16 33.08 &  86 & -2.12 & -0.89 & 8.58$\pm$0.24 & WR & 3 & NGC 5447 \\
1378-53061-023 & 0.00442 & +14 02 36.07 & +39 13 13.28 &  97 & -1.92 & -1.46 & 8.17$\pm$0.16 & WR & 2 & N/A  \\
1323-52797-002 & 0.00056 & +14 03 01.17 & +54 14 29.40 &  83 & -2.59 & -1.75 & 7.78$\pm$0.07 & WR & 3 & N/A  \\
1323-52797-014 & 0.00087 & +14 03 34.06 & +54 18 36.91 &  84 & -2.25 & -0.86 & 8.57$\pm$0.30 & WR & 3 & NGC 5461 \\
1325-52762-356 & 0.00085 & +14 03 39.84 & +54 18 56.87 &  88 & -2.29 & -0.67 & 8.75$\pm$0.23 & WR & 3 & NGC 5461 \\
1325-52762-350 & 0.00093 & +14 04 11.24 & +54 25 18.67 &  87 & -2.40 & -1.48 & 7.99$\pm$0.06 & Non-WR & 0 & N/A  \\
1642-53115-155 & 0.01194 & +14 04 14.87 & +36 43 32.67 & 113 & -2.14 & -0.88 & 8.54$\pm$0.29 & WR & 3 & MRK 1369 \\
1324-53088-234 & 0.00097 & +14 04 28.63 & +54 23 52.80 &  85 & -2.41 & -1.30 & 8.33$\pm$0.32 & Non-WR & 0 & N/A  \\
1325-52762-412 & 0.07731 & +14 09 56.76 & +54 56 48.89 &  89 & -2.00 & -1.32 & 8.34$\pm$0.05 & WR & 3 & SBS 1408+551A \\
2786-54540-084 & 0.00855 & +14 18 51.13 & +21 02 39.74 & 196 & -1.97 & -2.24 & 7.76$\pm$0.04 & Non-WR & 0 & N/A  \\
2746-54232-104 & 0.00771 & +14 23 48.53 & +14 38 16.54 & 194 & -2.04 & -0.48 & 9.04$\pm$0.33 & WR & 3 & N/A  \\
1381-53089-470 & 0.02230 & +14 26 28.17 & +38 22 58.67 &  98 & -1.87 & -1.96 & 7.77$\pm$0.07 & Non-WR & 0 & N/A  \\
1644-53144-564 & 0.08624 & +14 28 05.51 & +36 27 10.40 & 114 & -1.93 & -1.60 & 7.88$\pm$0.04 & Non-WR & 0 & N/A  \\
1827-53531-503 & 0.17350 & +14 29 47.01 & +06 43 34.97 & 133 & -1.80 & -1.17 & 8.79$\pm$0.04 & Non-WR & 0 & N/A  \\
0305-51613-604 & 0.01341 & +14 30 53.51 & +00 27 46.35 &   6 & -2.28 & -1.67 & 8.02$\pm$0.11 & WR & 1 & N/A  \\
2137-54206-310 & 0.01502 & +14 31 08.88 & +27 14 12.29 & 158 & -2.15 & -1.20 & 8.44$\pm$0.31 & Non-WR & 0 & MRK 0685 \\
1709-53533-215 & 0.00466 & +14 32 48.36 & +09 52 57.15 & 119 & -2.31 & -1.42 & 8.06$\pm$0.10 & WR & 2 & NGC 5669 \\
0920-52411-575 & 0.02740 & +14 48 05.38 & -01 10 57.72 &  61 & -2.02 & -1.73 & 7.98$\pm$0.07 & WR & 3 & SHOC 486 \\
1646-53498-616 & 0.00267 & +14 48 52.02 & +34 42 42.99 & 115 & -1.48 & -1.66 & 7.91$\pm$0.09 & Non-WR & 0 & UGC 09540 \\
1383-53116-110 & 0.00396 & +14 50 56.56 & +35 34 19.59 &  99 & -2.24 & -1.41 & 8.05$\pm$0.10 & WR & 1 & N/A  \\
2145-54212-388 & 0.08403 & +14 51 33.55 & +26 46 03.56 & 159 & -1.51 & -0.67 & 9.00$\pm$0.04 & WR & 1 & N/A  \\
1843-53816-087 & 0.00942 & +14 54 12.15 & +30 12 36.25 & 134 & -2.38 & -1.52 & 8.14$\pm$0.16 & Non-WR & 0 & N/A  \\
1844-54138-311 & 0.00610 & +14 56 36.63 & +30 13 52.36 & 135 & -2.31 & -1.50 & 8.01$\pm$0.08 & WR & 2 & N/A  \\
1399-53172-299 & 0.03255 & +15 09 34.18 & +37 31 46.11 & 100 & -2.01 & -2.09 & 7.81$\pm$0.04 & Non-WR & 0 & N/A  \\
2911-54631-344 & 0.04684 & +15 19 47.15 & +39 45 37.85 & 199 & -1.98 & -0.63 & 9.08$\pm$0.12 & WR & 2 & N/A  \\
1651-53442-255 & 0.06816 & +15 23 32.19 & +29 31 12.08 & 116 & -1.96 & -1.79 & 7.88$\pm$0.04 & Non-WR & 0 & N/A  \\
1679-53149-384 & 0.00826 & +15 26 30.31 & +41 17 22.34 & 118 & -1.83 & -1.04 & 8.42$\pm$0.38 & WR & 3 & UGC 09856 \\
2163-53823-546 & 0.03395 & +15 34 56.40 & +24 51 39.24 & 161 & -1.68 & -0.64 & 9.05$\pm$0.10 & WR & 2 & N/A  \\
0616-52442-364 & 0.00225 & +15 37 04.18 & +55 15 50.62 &  40 & -2.46 & -1.33 & 8.54$\pm$0.04 & Non-WR & 0 & N/A  \\
1725-54266-068 & 0.03772 & +15 45 43.55 & +08 58 01.35 & 120 & -1.69 & -2.01 & 7.72$\pm$0.05 & Non-WR & 0 & N/A  \\
2167-53889-071 & 0.01098 & +15 46 58.88 & +17 53 03.07 & 162 & -1.72 & -0.40 & 9.08$\pm$0.14 & WR & 3 & NGC 5996 \\
2524-54568-146 & 0.10704 & +16 06 27.54 & +13 55 47.88 & 184 & -1.99 & -1.28 & 8.76$\pm$0.06 & WR & 1 & N/A  \\
2527-54569-147 & 0.01217 & +16 15 17.02 & +13 01 33.08 & 185 & -1.96 & -1.57 & 7.96$\pm$0.07 & Non-WR & 0 & N/A  \\
0624-52377-361 & 0.00237 & +16 16 23.54 & +47 02 02.32 &  42 & -1.56 & -1.91 & 7.87$\pm$0.08 & WR & 3 & Arp 2 \\
0364-52000-187 & 0.03133 & +16 24 10.11 & -00 22 02.58 &  12 & -2.11 & -1.58 & 7.99$\pm$0.05 & WR & 3 & SHOC 536 \\
0624-52377-092 & 0.02993 & +16 26 04.26 & +46 22 05.79 &  41 & -1.36 & -0.39 & 9.25$\pm$0.03 & WR & 2 & N/A  \\
1570-53149-021 & 0.00910 & +16 47 10.66 & +21 05 14.51 & 109 & -1.83 & -2.06 & 7.74$\pm$0.07 & Non-WR & 0 & N/A  \\
1342-52793-112 & 0.03205 & +16 49 05.27 & +29 45 31.61 &  90 & -1.52 & -0.39 & 9.30$\pm$0.03 & WR & 1 & KUG 1647+298 \\
0976-52413-600 & 0.01195 & +17 12 36.63 & +32 16 33.42 &  67 & -1.99 & -1.96 & 7.79$\pm$0.06 & WR & 1 & N/A  \\
0978-52441-118 & 0.01483 & +17 18 53.45 & +30 11 36.20 &  68 & -1.92 & -0.84 & 8.60$\pm$0.27 & WR & 2 & IRAS 17169+3014 \\
0358-51818-504 & 0.04723 & +17 35 01.25 & +57 03 08.55 &  11 & -2.17 & -1.31 & 8.31$\pm$0.04 & WR & 2 & SHOC 579 \\
1115-52914-309 & 0.01381 & +20 47 59.21 & -00 10 53.98 &  74 & -2.18 & -1.17 & 8.44$\pm$0.34 & WR & 1 & N/A  \\
0673-52162-312 & 0.06669 & +22 25 10.13 & -00 11 52.84 &  46 & -1.85 & -1.81 & 7.90$\pm$0.04 & Non-WR & 0 & N/A  \\
1893-53239-476 & 0.02061 & +22 38 31.12 & +14 00 29.78 & 137 & -2.16 & -2.22 & 7.79$\pm$0.03 & Non-WR & 0 & N/A  \\
0742-52263-179 & 0.03029 & +23 01 23.59 & +13 33 14.79 &  50 & -2.06 & -1.68 & 7.90$\pm$0.04 & Non-WR & 0 & N/A  \\
0677-52606-533 & 0.03312 & +23 02 10.00 & +00 49 38.84 &  47 & -1.83 & -2.21 & 7.75$\pm$0.04 & Non-WR & 0 & N/A  \\
0650-52143-330 & 0.03592 & +23 56 21.96 & -09 04 07.42 &  43 & -1.51 & -1.43 & 8.00$\pm$0.07 & WR & 1 & N/A  \\
\hline 
\end{longtable}
\clearpage
\end{landscape}
\twocolumn

\end{document}